\documentclass[aps,prl,twocolumn,superscriptaddress,footinbib]{revtex4-2}  % ,linenumbers for review and submission
\usepackage{graphicx}  % needed for figures
\usepackage{dcolumn}   % needed for some tables
\usepackage{bm}        % for math
\usepackage{amssymb}   % for math
\usepackage{amsmath}   % for math
\usepackage{extarrows}
\usepackage{hyperref}
\usepackage{color} 

\newcommand{\bra}[1]{{\left\langle  #1 \right|}}
\newcommand{\ket}[1]{{\left|  #1 \right\rangle}}

\newcommand{\hide}[1]{}

\newcommand{\be}{\begin{equation}}
\newcommand{\ee}{\end{equation}}
\newcommand{\bes}{\begin{eqnarray}}
\newcommand{\ees}{\end{eqnarray}}
\newcommand{\tot}{\mathrm{tot}}
\newcommand{\sys}{\mathrm{S}}
\newcommand{\SB}{\mathrm{SB}}
\newcommand{\bath}{\mathrm{B}}

\newcommand{\e}[1]{{\mathrm{e}^{#1}}}

\begin{document}

\title{Global becomes local: Efficient many-body dynamics for global master equations}
\author{Alexander~Schnell} 
\email[Electronic address: ]{schnell@tu-berlin.de}
\affiliation{Technische Universit{\"a}t Berlin, Institut f{\"u}r Physik und Astronomie, 10623 Berlin, Germany}

\date{\today}

\begin{abstract}
This work makes progress on the issue of global- vs.\ local master equations. Global master equations like the Redfield master equation (following from standard Born- and Markov approximation) %suffer from the problem that they 
require a full diagonalization of the system Hamiltonian. This is especially challenging for interacting quantum many-body systems.
%, effectively preventing us from even ``writing down'' the concrete jump operators for large systems.
We discuss a  short-bath-correlation-time expansion in reciprocal (energy) space, leading to a series expansion of the jump operator, which avoids a diagonalization of the Hamiltonian. For a bath that is coupled locally to one site, this typically leads to an expansion of the global Redfield jump operator in terms of local operators. We additionally map the local Redfield master equation to a novel local Lindblad form, giving an equation which has the same conceptual advantages of traditional local Lindblad approaches, while being applicable in a much broader class of systems. %We demonstrate this by combining our approach with quantum trajectories, allowing us to find approximate solutions of the Redfield dynamics for extremely large Hilbert space dimensions. 
Our ideas give rise to a non-heuristic foundation of local master equations, which can be combined with established many-body methods. %like mean-field or matrix product approaches.
%Our ideas allow for applying methods like mean-field and matrix product ansatz to global master equations.
\end{abstract}

\maketitle

\emph{Introduction.}---
Foundational questions of the theory of open quantum systems have recently experienced a resurgence due to the high control and resolution of recent experiments on open quantum matter \cite{LabouvieEtAl16,LetscherEtAl17,HohmannEtAl17,HusmannEtAl18,ClarkEtAl20,BoutonEtAl20,OeztuerkEtAl21,NymanEtAl21,FontaineEtAl22,BlochEtAl22,Dogra1496,VrajitoareaEtAl22,RobertsEtAl23} and a new push from the theory side to study the dynamics of open quantum many-body systems \cite{CarusottoEtAl09,DiehlEtAl11,Prosen11,Prosen11_2,KarevskiEtAl13,VorbergEtAl13,LeviEtAl16,FischerEtAl16,LWu_2018,SorienteEtAl18,HestenEtAl18,WuEckardt19,XuEtAl19,froml2020ultracold,GluzaEtAl21,GladilinEtAl22,StitelyEtAl23}. While the limit of ultraweak system--bath coupling is well understood \cite{BreuerPetruccione,WeissBook,CarmichaelBook1}, leading to the quantum-optical master equation in Lindblad form, the underlying secular approximation is often violated by genuine quantum many-body systems where close degeneracies in the many-body spectrum are expected from exponentially large densities of states and vanishing finite-size gaps.  Hence one has to resort to finite system--bath coupling master equations where a multitude of master equations have been proposed \cite{HEOM89,BLaird1991,QUAPI95,CaoJCP,RCNazir14,TTM14,Imamoglu94,Garraway97,Mazzola09,JThingnaWJSheng2014,Nazir2018,SAlipour2019,NathanRudner20,Dvira21,Trushechkin21,PurkayasthaEtAl21,BeckerEtAl21,Trushechkin22,Cerisola2022,BeckerEtAl22,Winczewski21,Alicki22,LacerdaEtAl23,Shiraishi24}, each with their advantages and drawbacks \cite{HartmannStrunz20,Purkayastha22}. 

Nevertheless, the well-established Redfield master equation~\cite{BreuerPetruccione,WeissBook,CarmichaelBook1,AGRedfield65}  still often outperforms other approaches \cite{HartmannStrunz20,Purkayastha22,TupkaryEtAl23} (despite known problems regarding positivity violation and inaccurate steady states \cite{TMori2008,FlemingCummings2011,JThingaPHaenggi2012,ThingnaPRE13,TalknerHaenggi2020,CresserAnders21,Purkayastha22,TupkaryEtAl23,Anders2022}, which however can be cured efficiently \cite{BeckerEtAl22}\footnote{The master eq.\ of Ref.~\cite{BeckerEtAl22} is not of Lindblad form, so complete positivity is not guaranteed. Nevertheless positivity is typically preserved on second order.}). It belongs to the class of global master equations \cite{LandiEtAl22}, meaning that it involves the global eigenstates and -energies of the system. This is challenging for extended quantum many-body systems \cite{NathanRudner20,Shiraishi24,Garcia2024} especially far from equilibrium, where an effective low-energy theory in terms of free quasiparticles is not possible.   Also it prevents one from directly applying established many-body methods \cite{XuEtAl19}. Alternatively, one therefore often resorts to local Lindblad equations \cite{LandiEtAl22}, where for site-locally coupled baths one introduces local Lindblad jump operators \cite{HWichterich2007,Prosen11,Prosen11_2,KarevskiEtAl13,LandiEtAl14,SchuabEtAl16,ZnidaricEtAl17,LandiEtAl22,PrechEtAl23}. 
Nevertheless, such approaches often violate consistency with statistical mechanics \cite{Purkayastha22}. Additionally, for out-of-equilibrium systems, e.g.~coupled to multiple baths, and strongly coupled systems, the specific details of the environment and how the system is coupled to it influence the  steady state \cite{ThingnaPRE13,LetscherEtAl17,SchnellEtAl17,BeckerEtAl22}, so that one needs microscopically derived equations.
Here we show an alternative scheme that maps the Redfield master equation into a local form, which on leading order can be further approximated to Lindblad form with local jump operators.

\begin{figure}
\centering
\includegraphics[scale=0.82]{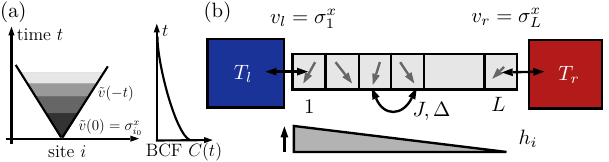}
\caption{(a)~Sketch of Redfield jump operator $u$. In the interaction picture, an initially local coupling operator $\tilde v(0)$ is spreading in space. This operator is convoluted with the bath-correlation function (BCF) $C(t)$. 
%(b)~Convergence of the Taylor series around $\varepsilon_0 = -2J$ (top), $2J$ (bottom) for $W(E)$ of a pure ohmic bath at temperature $T=J$ (solid blue line). 
(b)~System under investigation: An XXZ chain is connected at both ends to heat baths of temperature $T_l$, $T_r$ with coupling operators $v_l$, $v_r$. Additionally, an external field gradient $h_i$ in the $z$-direction is applied.}
\label{fig:sketch-bcf}
\end{figure}

\emph{Redfield master equation.}---
Consider a typical open quantum system setup with total Hamiltonian
$H_\tot = H_\sys + H_\SB + H_\bath$, where $H_\sys$ and $H_\bath$ are Hamiltonians that act on the system and bath Hilbert space only and $H_\SB$ is the system--bath interaction Hamiltonian, which for simplicity we assume is a direct product $H_\SB = v \otimes B$ of system- and bath coupling operators $v$ and $B$ respectively. The generalization to arbitrary interaction Hamiltonians is discussed in the Supplemental Material (SM) \cite{SM}. Performing Born- and Markov approximations, the dynamics of the reduced density matrix of the system $\varrho(t)=\mathrm{Tr}_\bath \varrho_\tot(t)$, is given by the Redfield quantum master equation~\cite{BreuerPetruccione,WeissBook,CarmichaelBook1,AGRedfield65,BeckerEtAl22,Thingna2013PHD,Schnell2019PHD}
\begin{equation}
	\partial_t \varrho(t) = -i \left[ H_\sys, \varrho(t) \right] +\left( \left[ u \varrho(t), v \right] + \mathrm{h.c.}\right).
	\label{eq:Redfield}
\end{equation}
As sketched in Fig.~\ref{fig:sketch-bcf}(a), the operator
%\begin{equation}
$	u = \int_0^\infty \mathrm{d}\tau  \tilde{v}(-\tau) C(\tau)$  %\cite{BeckerEtAl22,Thingna2013PHD,Schnell2019PHD}
%	\label{eq:uoperator}
%\end{equation}
convolutes the coupling operator in the interaction picture $\tilde v(t) = \e{iH_\sys t} v   \e{-iH_\sys t}$ with the bath-correlation function $C(t) = \langle \tilde B (t) \tilde B(0) \rangle_\bath$, where $\tilde B(t) = \e{iH_\bath t} B  \e{-iH_\bath t}$. Throughout the manuscript we set $\hbar = k_\mathrm{B}=1$.

While the Redfield eq.~\eqref{eq:Redfield} generally leads to more accurate results when compared to heuristic Lindblad descriptions (like local Lindblad equations), it has two major deficiencies that prohibit its straight forward application to genuine quantum many-body systems: 
1) Since it does not obey Lindblad structure, its dynamics will generally violate positivity (this issue can be mitigated  by making connections to statistical mechanics \cite{BeckerEtAl22}).
%Positivity violations are especially common in interacting many-body systems where the energy spectrum is unbounded, so thermal occupations are extremely close to zero and small errors directly lead to positivity violation. 
And 2) to implement it,
one generally introduces the system's eigenbasis $H_\sys = \sum_k E_k \ket{k}\bra{k}$, giving 
%\begin{equation}
	 %= \int_0^\infty \mathrm{d}\tau \sum_{kq} e^{-i(E_k-E_q) \tau} \hat{v}_{kq} C(\tau)  
	$u = \sum_{kq}  \hat{v}_{kq} W(E_k-E_q)$ \cite{SM}
%\end{equation}
with eigenspace-projected coupling operator $ \hat{v}_{kq} = \ket{k}\bra{k}v\ket{q}\bra{q}$ and Fourier-Laplace transform $W(E)=\int_0^\infty \mathrm{d}\tau  \mathrm{e}^{-i E \tau} C(\tau)$ of the bath correlation function. This is problematic since it requires a full diagonalization of the many-body Hamiltonian which is generally a hard task. And even then, if the coupling operator $v$ can be represented as a sparse matrix (e.g.~if $v$ is an operator that is local in space) the matrix $u$ is generally a dense matrix, making the numerical solution of the equation memory- and time consuming. 
We therefore call the Redfield eq.~\eqref{eq:Redfield} a global master equation since it involves the global eigenstates of the system.
Put differently, this practically hinders us from even `writing down' the global Redfield master equation in a way that we can combine it with standard many-body methods.
% In the following, we propose a way around this problem by performing an expansion of the function $W(E)$ around a given transition energy $E= \varepsilon_0$.
Note that, different from here, the term `global master equation' is sometimes used specifically  for the quantum optical master equation \cite{BreuerPetruccione,WeissBook,CarmichaelBook1}, a Lindblad master equation that follows from the Redfield equation after rotating-wave approximation. %, the so-called quantum optical master equation. 

\emph{Correlation-time expansion}---To write the Redfield eq.~\eqref{eq:Redfield} in  local form, we use ideas similar to derivations of quantum Brownian motion \cite{BreuerPetruccione} where 
one expands $\tilde v(-\tau)$ in first order for short times $\tau$, which is valid for short bath-correlation times, cf.~Fig.~\ref{fig:sketch-bcf}(a).
Here we expand to arbitrary order  \cite{NathanRudner20} and additionally insert $1=\mathrm{e}^{-i(E-\varepsilon_0) \tau}\vert_{E=\varepsilon_0}$  to  rewrite 
\begin{align}
	u &= \int_0^\infty \mathrm{d}\tau \mathrm{e}^{- i ([H_\sys, \cdot] -\varepsilon_0 \cdot)\tau} [v] \  \mathrm{e}^{-iE \tau}\vert_{E=\varepsilon_0} C(\tau),
	\label{eq:u-exact}
\end{align}
where in the exponential it enters a superoperator acting as $([H_\sys, \cdot] -\varepsilon_0 \cdot)[v]=[H_\sys, v] -\varepsilon_0 v$.
By expanding the first exponential in a series in $\tau$, which is valid as long as $C(\tau)$ decays on a bath-correlation timescale that is short against the timescale of operator spreading of $\tilde{v}(-\tau)$ (typically confined by Lieb-Robinson bounds \cite{LiebRobinson72,Kliesch2014,Shiraishi24}, cf.\ Fig.~\ref{fig:sketch-bcf}(a)), then using $-i\tau =\frac{\partial}{\partial E}$ and exchanging the integral and the derivative, we find a Taylor-like expansion, 
\begin{align}
	u %&= e^{ [H_\sys, \cdot] \frac{\partial}{\partial E}} [v] \  W(E) \vert_{E=0} \\
	%&= W_{E_0} v + W'_{E_0}  \left(\left[H_\sys, v\right]  -E_0 v \right)+ \frac{W''_{E_0}}{2} \left[H_\sys, \left[H_\sys, v\right] -E_0 v \right]  + ...
	%&\simeq u^{(N)} = W(\varepsilon_0) v + \sum_{n=1}^{N} \frac{W^{(n)}(\varepsilon_0)}{n!}([H_\sys, \cdot] -\varepsilon_0 \cdot)^n [v] %\left[H_\sys, v\right]^{(n)}	
	%&\simeq u^{(N)} 
	&\approx \sum_{n=0}^{N} \frac{W^{(n)}(\varepsilon_0)}{n!}([H_\sys, \cdot] -\varepsilon_0 \cdot)^n [v] \equiv \mathcal{T}^{\varepsilon_0}_N[v],
	\label{eq:u-approx}
\end{align}
where we call the corresponding master equation the \emph{ad-hoc local Redfield equation}.
It enters the $n$th derivative $W^{(n)}(\varepsilon_0)$ % $W^{(n)}_{\varepsilon_0} =W^{(n)}(\varepsilon_0)$ 
and $([H_\sys, \cdot] -\varepsilon_0 \cdot)^n$ denotes $n$ times  the application of the superoperator. 
We observe that the terms in Eq.~\eqref{eq:u-approx} represent an expansion in small parameter $E_k-E_q - \varepsilon_0$, which is the deviation of the transition energy $E_k-E_q$ (due to the commutator structure) from the expansion energy $\varepsilon_0$ of a thought quantum jump from $q$ to $k$ that is induced by $u$.
%As we show in the  supplemental material (SM) \cite{SM}, in energy space, this can  be understood as an expansion of the Redfield jump operator $u$ around $E=\varepsilon_0$, i.e.~at transition energies $E$ that are close to $\varepsilon_0$ (see Fig.~\ref{fig:sketch-bcf}(b)). We also formally prove \cite{SM} that the series in Eq.~\eqref{eq:u-approx} converges as long as all relevant transition energies are in the convergence radius of the scalar Taylor series around $\varepsilon_0$. %for two typical Taylor series around different $\varepsilon_0$). 
Eq.~\eqref{eq:u-approx} has the significant advantage that if the expansion is truncated, a \emph{diagonalization} of the system Hamiltonian is \emph{not required}, which is a remarkable simplification in the case of interacting many-body systems.
Also if $v$ and $H_\sys$ have a sparse representation, the nested commutators remain sparse. In the case where $v$ is a site-local operator and $H_\sys$ involves nearest-neighbor terms only, 
Eq.~\eqref{eq:u-approx} can be understood as an expansion in local operators, exactly reflecting our intuition from Fig.~\ref{fig:sketch-bcf}(a). 

%To understand what the transition energies must be small compared to, 
Let us consider a pure ohmic bath, $H_\mathrm{B}= \sum_{\alpha} \omega_\alpha b^\dagger_\alpha b_\alpha $, $B =\sum_{\alpha} c_\alpha (b_\alpha^\dagger + b_\alpha)$, with a continuum of modes $\alpha$ and spectral density $J(E)= \sum_\alpha c_\alpha^2 [\delta(E-\omega_\alpha)-\delta(E+\omega_\alpha)]=\gamma E$ with dissipation strength $\gamma$. %where $C(t) = \int_{-\infty}^\infty \mathrm{d}E J(E) [e^{iEt} n_\mathrm{B}(E)+e^{-iEt} (n_\mathrm{B}(E)+1)$.
In this case $W(E)$ has no imaginary part and takes the form  $W(E)=J(E)/[\e{E/T}-1]$, where $T$ is the temperature of the bath. %Due to these vanishing imaginaries, the Redfield master equation gives rise to a valid description of the reduced dynamics at second order of the system--bath coupling at all times, since the correction term that was found in Ref.~\cite{BeckerEtAl22} vanishes.
%Expanding, e.g., around $\varepsilon_0=0$, we find $W(0)=\gamma T$, $W'(0)=-\gamma/2$, $W''(0)=\gamma/(6T)$. 
Calculating the derivatives $W^{(n)}(\varepsilon_0)$, we observe that each order involves higher terms in $1/T$. Hence, Eq.~\eqref{eq:u-approx} can be regarded as a high-temperature expansion of the Redfield jump operator.
At high temperatures, when compared to the typical transition energies in the system, as we show in the  SM \cite{SM}, this expansion energy is in principle arbitrary.
This is also confirmed by our intuition that at high temperatures $C(t)$ decays  rapidly and for $T\to \infty$ tends to a delta function, giving the leading $N=0$ term, in Eq.~\eqref{eq:u-approx}. Also note that by setting $N=1$ and $\varepsilon_0 =0$, we recover the standard result for quantum Brownian motion \cite{BreuerPetruccione}. At intermediate temperatures however, and if the bath is coupled locally to the many-body system, we can use additional information about the coupling operator $v$.

%To generalize our results also to other quantum systems,
\emph{Optimal expansion energy}---What remains is to specify the optimal expansion energy $\varepsilon_0$.
As it turns out, for coupling operators $v$ that are site local, an optimal choice can be found from introducing \emph{local} transition energies at the coupling site.
We thus assume a generic quantum many-body system, whose Hamiltonian can typically be understood as the sum of onsite terms 
and the residual interaction- or tunneling terms, i.e.~$H_\sys =  \sum_{i=1}^{L} H_\sys^{\mathrm{loc}, i} + H_\sys^{\mathrm{resid}}$.
Here $H_\sys^{\mathrm{loc}, i}=I_1 \otimes \cdots \mathcal{H}_i \otimes \cdots I_L$ only acts nontrivially on site $i$ of a many-body system with  $L$ sites, and $I_j$ is the identity on site $j$. 
%\begin{align}
%\label{eq:H-gener}
% \end{align}
\begin{figure}
\centering
\includegraphics[scale=0.8]{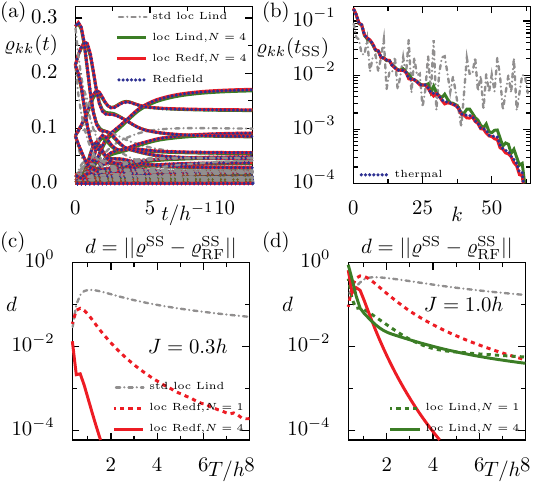}
\caption{(a) Population dynamics of the many-body eigenstates $k$ of the boundary-driven XXZ chain with $T_l=T_r = T=2.2h$ and $J=h$. Exact Redfield- (dotted blue), local Redfield- (solid red), local Lindblad- (solid green), both  $N=4$, and standard local Lindblad (dash-dotted) dynamics. Coherences in the SM \cite{SM}. (b) Steady-state populations for the parameters and methods of (a).
(c),(d) Trace distance $d$ of  the steady state of the local Redfield- (red), local Lindblad (green)  to the exact Redfield eq.~as a function of temperature $T$ for  orders $N=1$ (dashed), $N=4$ (solid) and (c) $J=0.3h$, (d) $J=h$, as well as the standard local Lindblad (dash-dotted).
%(d)~As in (c) but for the approximate Lindblad form in Eq.~\eqref{eq:Redf-aslind}. 
Other parameters $\gamma=0.25 h, L=6, \Delta = 0.7, \delta=-0.07h$, initial state is pure $\ket{\psi(0)}=[ (\ket{\uparrow}+\ket{\downarrow})/\sqrt{2}]^{\otimes L}$.}
\label{fig:error-rf}
\end{figure}
Even though exact diagonalisation is generally a hard task for interacting many-body systems, the site-local Hamiltonian can typically be diagonalized straightforwardly
%\begin{align}
$H_\sys^{\mathrm{loc}, i} \ket{k^{i}}= E^{\mathrm{loc},i}_k \ket{k^{i}}$.
%$\mathcal{H}_i\ket{k^{i}}= \mathcal{E}^{i}_k \ket{k^{i}}$.
% \end{align}

To derive the local Redfield master equation, we assume that the coupling operator $v=I_1 \otimes \cdots \mathcal{V}_{i_0} \otimes \cdots I_L$ is acting site-locally at site $i_0$ only.
We then perform a spectral decomposition of the operator $v$ with respect to this local on-site basis
%\begin{align}
$v = \sum_{kq} \hat{v}^{i_0}_{kq}$, with $\hat{v}^{i_0}_{kq}= \ket{k^{i_0}}\bra{k^{i_0}}v \ket{q^{i_0}}\bra{q^{i_0}}$.
% \end{align}
 Then, using the arguments of Eq.~\eqref{eq:u-approx} at each of the terms in the sum, %$u \approx \sum_{kq} \mathcal{T}^{\varepsilon^{i_0}_{kq}}_N[\hat{v}^{i_0}_{kq}] $,
  we finally arrive at the  \emph{local Redfield equation}
 \begin{align}
u \approx %\sum_{kq} \mathcal{T}^{\varepsilon^{i_0}_{kq}}_N[\hat{v}^{i_0}_{kq}] =
 u_{\mathrm{loc}}= \sum_{kq} \sum_{n=0}^{N} \frac{W^{(n)}(\varepsilon^{i_0}_{kq})}{n!} ([H_\mathrm{S}, \cdot ] -\varepsilon^{i_0}_{kq} \cdot )^n [\hat{v}^{i_0}_{kq}],
\label{eq:locRedf}
 \end{align}
 where we have used the freedom to choose different Taylor expansion energies  $\varepsilon^{i_0}_{kq}$ for each term of the spectral decomposition.
 %$\varepsilon^{i_0}_{kq}=E^{\mathrm{loc},i_0}_k -E^{\mathrm{loc},i_0}_q$.
 %for each of the 
 %spectrally decomposed jump operators $\hat{v}^{i_0}_{kq}$. 
%we have by definition
% \begin{align}
%u \overset{H_\sys^{\mathrm{resid}}=0}{=} \sum_{kq}  W(E^{\mathrm{loc},i_0}_k -E^{\mathrm{loc},i_0}_q)  \hat{v}^{i_0}_{kq},
%\label{u-ex-local}
 %\end{align}
% without approximation. In the same limit it holds that 
 % \begin{align}
%[H_\mathrm{S}, \hat{v}^{i_0}_{kq} ] \overset{H_\sys^{\mathrm{resid}}=0}{=} (E^{\mathrm{loc},i_0}_k -E^{\mathrm{loc},i_0}_q) \,  \hat{v}^{i_0}_{kq}.
% \end{align}
% Hence by setting the expansion energies to these local transition energies, $\varepsilon^{i_0}_{kq}=E^{\mathrm{loc},i_0}_k -E^{\mathrm{loc},i_0}_q$,
 Note that in the limit of vanishing  $H_\sys^{\mathrm{resid}}=0$, we observe that we can achieve $([H_\mathrm{S}, \cdot ] -\varepsilon^{i_0}_{kq} \cdot ) [\hat{v}^{i_0}_{kq}]=0$ simply  by setting the expansion energies to the local transition energies $\varepsilon^{i_0}_{kq}=E^{\mathrm{loc},i_0}_k -E^{\mathrm{loc},i_0}_q$.
 Thus, for $H_\sys^{\mathrm{resid}}=0$, we
 reproduce the \emph{exact} operator $u$ at all temperatures $T$ already in the lowest order $N=0$ and all higher orders vanish. %Plugging this choice for the transition energies in Eq.~\eqref{eq:locRedf} gives  the local Redfield equation for a general quantum system. 
 Thus, we expect good convergence  as long as $H_\sys^{\mathrm{resid}} \ll H_\sys$.
 %We expect that, as long as $H_\sys^{\mathrm{resid}} \ll H_\sys$, the local Redfield  in Eq.~\eqref{eq:locRedf} converges much faster than the ad-hoc local Redfield in Eq.~\eqref{eq:u-approx} \cite{SM}.
 
 To illustrate this, we 
%In case of the XXZ Hamiltonian of the main text, e.g. this decompositon reads
study an XXZ model with external field $h_i$ in the $z$-direction,
%  \begin{align}
%H_\sys ={ -J \sum_{i=1}^{L-1}\left( \sigma^x_{i} \sigma^x_{i+1} +  \sigma^y_{i} \sigma^y_{i+1} + \Delta  \sigma^z_{i} \sigma^z_{i+1}\right)} + {\sum_{i=1}^{L} h_i \sigma^z_{i}}%_{H_\sys^0},
%\label{eq:H-xxz}
 %\end{align}
   \begin{align}
H_\sys ={ -J \sum_{i=1}^{L-1}\left( \sigma^x_{i} \sigma^x_{i+1} +  \sigma^y_{i} \sigma^y_{i+1} + \Delta  \sigma^z_{i} \sigma^z_{i+1}\right)} + \sum_{i=1}^{L} H_\sys^{\mathrm{loc}, i},
%\label{eq:H-xxz-supp}
\label{eq:H-xxz}
 \end{align}
with $H_\sys^{\mathrm{loc}, i}=h_i \sigma^z_{i}$, where the field $h_i=h+(i-1) \delta$ can be tuned to have a gradient for $\delta\neq 0$, cf.~Fig.~\ref{fig:sketch-bcf}(b). 
 We couple the spin system to two pure ohmic baths, with temperature $T_l$ ($T_r$) to the leftmost (rightmost) site operator $v_l = \sigma^x_{1}$ ($v_r = \sigma^x_{L}$). Here, the dissipator %in the Redfield eq.~\eqref{eq:Redfield} and the approximate Lindblad form \eqref{eq:Redf-aslind} 
 is given as the sum %of the coupling to the individual baths 
 $\partial_t \varrho = -i \left[ H_\sys, \varrho \right] +\sum_{j=l,r} \left( \left[ u_j \varrho, v_j \right] + \mathrm{h.c.}\right)$.
In this case, the decomposition in terms of  the local eigenstates $\ket{\downarrow^1}, \ket{\uparrow^1}$ of $\sigma^z_1$  reads  $v_l %= \ket{\downarrow^1}\bra{\uparrow^1}+\ket{\uparrow^1}\bra{\downarrow^1}\equiv
=\sigma_1^- + \sigma_1^+$, with $ \sigma_i^\pm = ( \sigma_i^x \pm i\sigma_i^y)/2$. We find the local Redfield $u_l\approx \mathcal{T}^{\varepsilon_-^{1}}_N[\sigma_1^-]+ \mathcal{T}^{\varepsilon_+^{1}}_N[\sigma_1^+]$ %. For  $J=0$, we find $[H_\sys, \sigma_1^\pm] = \pm 2 h_1  \sigma_1^\pm$, so by setting the expansion energies to 
with local transition energies $\varepsilon^{1}_\pm=\pm 2 h_1$. %, the expansions terminate already after the leading term $N=0$, and we reproduce the exact Redfield operator at \emph{all} temperatures $T$. Thus, with this choice we expect good performance also for $J\ll h$. This equation, with this choice for $\varepsilon_\pm$ is what we call the \emph{local Redfield equation}. The generalization to the right bath and other  quantum systems is straight forward and outlined in the SM \cite{SM}.

\begin{figure}
\centering
\includegraphics[scale=0.8]{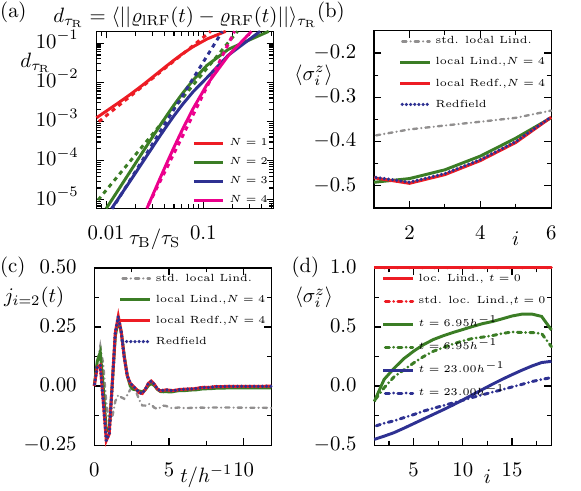}
\caption{%(a)~%Steady-state distance $d$ for our (solid, dashed) and the standard local Lindblad eq.~(dash-dotted). 
(a)~Error of the local Redfield averaged over the initial relaxation time $\tau_\mathrm{R}$ vs.~bath memory time $\tau_\mathrm{B}$ in units of operator-spreading time $\tau_\mathrm{S}$ for different orders $N$. Dashed lines are power laws $\propto (\tau_\mathrm{B}/\tau_\mathrm{S})^{N+1}$.
(b)~Steady-state average magnetization $\langle \sigma_i^{z} \rangle$ and (c)~local magnetization current $j_{i=2}(t)$ at site $i=2$ as function of time $t$ for methods and parameters of Fig.~\ref{fig:error-rf}(a), i.e.~$L=6$. (d)~Snapshots of the magnetization profile $\langle \sigma_i^{z} \rangle$ for our new, $N=1$, (solid) and the traditional local Lindblad (dash-dotted) equation for chain length $L=19$ at three different times (red, green, blue). Results obtained using quantum trajectory unravelings with $N_\mathrm{traj}=5000$ trajectories. Other parameters as in Fig.~\ref{fig:error-rf}(a), but  $\ket{\psi(0)} = \ket{\uparrow, \dots, \uparrow }$.}
\label{fig:locallind}
\end{figure}

In Fig.~\ref{fig:error-rf}(a) we plot the typical relaxation dynamics of the XXZ spin chain. The dotted blue lines show the  dynamics of the many-body eigenstates under the exact Redfield master equation for an intermediate temperature $T_l=T_r =T=2.2h$ and  interaction strength $J=h$. %with nearest-neighbor interaction constant $J$. 
The solid red lines show the dynamics under the local Redfield equation of order $N=4$. We observe exceptionally close agreement both throughout the complete relaxation dynamics and the steady state, Fig.~\ref{fig:error-rf}(b), where even the correct thermal distribution (which here is the steady state of the exact Redfield eq.) is reproduced. Therefore all observables, e.g.~entropy production, are captured with high accuracy.  Fig.~\ref{fig:error-rf}(c),(d) show  the trace distance, $\vert \vert A \vert \vert =\mathrm{Tr} \sqrt{{A^\dag A}}$, of the steady state of the local Redfield at order $N=1$ (dashed red), $N=4$ (solid red) to the exact Redfield eq.~at interaction strength (c) $J=0.3h$ and (d) $J=h$. %Other truncation orders $N$ are shown in the SM \cite{SM}. 
%We observe that, as expected, the expansion of the operator $u$ converges well at temperatures $T \gg J$.
%This is remarkable since the spectral width of the many-body spectrum at the given parameters is $18J$, which means that also possible high-energy transitions are well captured by our appproach.
% An intuitive explanation on why higher values of  $\varepsilon_0$ perform better, can be found from Fig.~\ref{fig:sketch-bcf}(b): If $\varepsilon_0 = -2J$ (top panel) the linear term $N=1$ in the Taylor series leads to negative rates $W(E)$ at transition energies $E \gtrsim 0.5J$, which can cause the Redfield dynamics to become unstable. For  $\varepsilon_0 = 2J$ (bottom panel) such negativities only occur at much higher energies $E\gtrsim 2.5J$ which are unlikely at $T=J$.
Note that the error plots in Fig.~\ref{fig:error-rf}(c),(d) qualitatively do not change by increasing the system size $L$. 
In order to validate our intuition about the short bath-memory time expansion, in Fig.~\ref{fig:locallind}(a) we plot
the  trace distance $d_{\tau_\mathrm{R}}$ averaged over the  initial relaxation dynamics up to time $\tau_\mathrm{R} =\gamma^{-1}$ as a function of bath memory time $\tau_\mathrm{B} =(2\pi T)^{-1}$ \cite{SM} relative to the time scale $\tau_\mathrm{S}=J^{-1}$ of operator spreading the system.  This distance follows the dashed power laws $(\tau_\mathrm{B}/\tau_\mathrm{S})^{N+1}$, which is the expected error scaling of our short-time expansion up to order $N$ in Eq.~\eqref{eq:u-exact}.
%Hence, for a given bath model, the expansion parameter $\varepsilon_0$ can be optimized at small system sizes and can then be used to study large systems.
Details are discussed in the SM \cite{SM}, where we %additionally show the approximate Redfield dynamics for a different bath model, a Lorentz-Drude bath. %with a spectral density $J(E)$ that is ohmic with a cutoff.
% For such a bath, the imaginary part of $W(E)$ is nonzero. 
%We
also show similar good performance  for other bath models, e.g.~a Lorentz-Drude bath,
as well as other quantum many-body systems, a Bose-Hubbard chain coupled to a thermal particle reservoir.

\emph{Local Lindblad form}--- The analytical form of $u$ also allows us to approximate the dynamics with a Lindblad form. %, which allows us to unravel the dynamics with standard quantum-trajectory algorithms  \cite{KMolmer1993,GardinerEtAl92,ADaley2014,LandiEtAl22}. 
For our purposes, the standard rotating-wave approximation is unfeasible since it relies on the knowledge of the full many-body spectrum.
Instead, we follow Ref.~\cite{BeckerEtAl21} and first rewrite Eq.~\eqref{eq:Redfield} in the form
%\begin{align}
	$\partial_t \varrho %= -i \left[ H_\sys-\frac{i}{2}(vu-u^\dagger v), \varrho \right] + \left( u\varrho v - \frac{1}{2} \lbrace vu,\varrho \rbrace + \mathrm{h.c.}\right)\\
	= -i \left[ H_\mathrm{eff}, \varrho \right] + \mathcal{D}[\varrho],$
%\end{align}
with effective Hamiltonian $H_\mathrm{eff} = H_\sys-\frac{i}{2}(vu-u^\dagger v)$ and dissipator
%\begin{align}
%	$\mathcal{D}[\varrho] =\left( u\varrho v - \frac{1}{2} \lbrace vu,\varrho \rbrace + \mathrm{h.c.}\right)$.
%\end{align}
%We observe that the dissipator %already has a structure of a pseudo-Lindblad form,
%has the structure
%\begin{align}
%	$\mathcal{D}[\varrho] =W_0 \sum_{ij}  \sigma^x_{ij} \left( A_i\varrho A_j^\dagger - \frac{1}{2} \lbrace A_j^\dagger A_i,\varrho\rbrace\right)$,
%\end{align}
%where we set $A_1 = v, A_2 = u/W_0$ 
%By diagonalizing the $\sigma^x$ Pauli matrix, we find the pseudo-Lindblad form,
%\begin{align}
	$\mathcal{D}[\varrho] =  D_{L_+}[\varrho] - D_{L_-}[\varrho]$ with  $D_{L}[\varrho]=\left( L\varrho L^\dagger - \frac{1}{2} \lbrace L^\dagger L,\varrho \rbrace\right)$.
%\end{align}
It enter jump operators
%\begin{align}
%	$L_\pm = \left({u}\pm {W({\varepsilon_0})}  v \right) /  \sqrt{2W({\varepsilon_0})}$, where we assume that $W({\varepsilon_0})$ is real (a derivation and the expressions for complex $W({\varepsilon_0})$ is given in the SM \cite{SM}) and nonzero, $W({\varepsilon_0}) \neq 0$.
	$L_\pm = \left({u}\pm {\lambda}  v \right) /  \sqrt{2\lambda}$, where there occurs a free parameter $\lambda \in \mathbb{R}$   \cite{SM,BeckerEtAl21,BeckerEtAl23}. 
%(expressions for complex $W(E)$ and $\lambda$ are given in the SM \cite{SM}).
%where we assume that $W({\varepsilon_0})$ is real (a derivation and the expressions for complex $W({\varepsilon_0})$ is given in the SM \cite{SM}) and nonzero, $W({\varepsilon_0}) \neq 0$.
%\end{align}
Note that  the second term in the Redfield dissipator $\mathcal{D}$ %occurs a \emph{negative} rate 
has a negative `rate'
which is why it violates Lindblad form. 
However, %by using the expansion in 
for the ad-hoc local Redfield eq.~\eqref{eq:u-approx}, we observe that by setting the free parameter to $\lambda = W(\varepsilon_0)$, $L_-$ vanishes in leading order $N=0$ and thus can be neglected at high temperatures \cite{SM}. %, and since it appears twice in the dissipator, by neglecting it, we omit terms of order $(\Delta/T)^2$. 
%This idea, however without the analytical insight into the operator $u$,  has already been discussed in Ref.~\cite{BeckerEtAl21}.
Hence, the expansion presented here also sheds light on the recently proposed truncation~\cite{BeckerEtAl21}.
 
 %Finally, %for low transition energies, %by omitting terms of order $(\Delta/T)^2$ and higher, 
For the local Redfield eq.~\eqref{eq:locRedf}, an optimal parameter choice can be found 
%by minimizing the norm of $L_-$ acting on a local equilibrium state, giving %$\lambda = \vert\vert u \vert\vert$
by averaging the above optimal value with weights according to local thermal equilibrium \cite{SM}, 
%$\lambda_{j}= \sqrt{\sum_{s=\pm}W(\varepsilon^{j}_{s})^2 e^{\varepsilon^{j}_{s}/T}}/ \sqrt{\sum_{s=\pm} e^{\varepsilon^{j}_{s}/T}}$ with $j=l,r$. 
%The generalization to other quantum systems and bath models is given in the SM \cite{SM}.
$\lambda_\mathrm{loc}= \sqrt{ \sum_{kq}  W(\varepsilon^{i_0}_{kq})^2 r_{kq}^2}/ \sqrt{\sum_{kq} r_{kq}^2}$  with $ r_{kq} = \bra{k^{i_0}}v \ket{q^{i_0}} e^{-E^{\mathrm{loc},i_0}_q/T}$.
The local Redfield eq.~can therefore approximately be rewritten in Lindblad form
% \begin{align}
$	\partial_t \varrho 
	\approx -i \left[ H_\mathrm{eff}, \varrho \right] + D_{L_{+}}[\varrho], $ %+ D_{L_+}[\varrho]. %+L\varrho L^\dagger - \frac{1}{2} \lbrace L^\dagger L,\varrho \rbrace,
%	\label{eq:Redf-aslind}
%\end{align}
 %with jump operator  $L_+= \left(2 {W_0}  v + W_0' [H_\sys, v] \right) /  \sqrt{2W_0}$.
 %\begin{align}
%$ L_+= \frac{1}{\sqrt{2 W_0}} \left(W_0 v +\sum_{n=0}^{\infty} \frac{W^{(n)}_0}{n!} \left[H_\sys, v\right]^{(n)} \right) $.
 %\end{align}
 %and $n$th order nested commutator  $\left[H_\sys, v\right]^{(n+1)} =[H_\sys, \left[H_\sys, v\right]^{(n)}] $, with $\left[H_\sys, v\right]^{(0)}=v$.
 with $L_{+} = \left({u}_\mathrm{loc}+ {\lambda_\mathrm{loc}}  v \right) /  \sqrt{2 \lambda_\mathrm{loc}}$ and the local approximation $u_\mathrm{loc}$, Eq.~\eqref{eq:locRedf}.
We call this our \emph{local Lindblad equation}. 
As we show in Fig.~\ref{fig:error-rf}(a),(b), it predicts the dynamics as well as the steady state quite accurately at intermediate and high temperatures.
 We plot the corresponding error in green in Fig.~\ref{fig:error-rf}(d). Surprisingly, for $N=1$, by neglecting the negative jump operator, the error decreases when compared to the local Redfield equation at the same order $N=1$.
Nevertheless, by increasing the order to $N=4$ the accuracy does only increase slightly and there is no convergence towards the exact Redfield result, due to the residual error that we make by neglecting the negative jump operator.
 This local Lindblad form is highly desired because it allows for stochastic unraveling with standard quantum trajectory algorithms \cite{KMolmer1993,GardinerEtAl92,ADaley2014,LandiEtAl22}. Since %when cutting off the expansion of $u$ on order $n$, 
 the resulting jump operator $L_+$ %\approx\left({u}^{(N)}+ {W({\varepsilon_0})}  v \right) /  \sqrt{2W({\varepsilon_0})}$ 
 is sparse and analytically obtainable, 
it is of great advantage over the full-rank exact many-body operator $u$, or the Lindblad jump operators that follow after rotating-wave approximation. %(that has to be obtained by ED). 
Both of these forms with local jump operators also allow for combining with other approaches like mean-field equations or tensor network methods \cite{ADaley2014}.
 Also note that there exist quantum trajectory unravelings directly of the Redfield equation \cite{breuerpetruccione1999,KondovEtAl03,breuer2004,DonvilMuratore22,BeckerEtAl23}. %, which can easily be combined with our local Redfield equation.
%however they typically require  a larger number of trajectories $N_\mathrm{traj}$ than the unraveling of the Lindblad equation. 

 \emph{Benchmark against traditional local Lindblad}---Our local Lindblad equation provides an alternative to the widely used traditional local Lindblad equations while, at the same time, not suffering from typical problems of the standard local Lindblad equations \cite{LandiEtAl22}. 
Using a  standard local Lindblad approach yields the master equation 
%   \begin{align}
$\partial_t \varrho 
	= -i \left[ H_\sys, \varrho \right] + \sum_{i=1, L} \left( \gamma_i^- D_{\sigma^{-}_i}[\varrho]+ \gamma_i^+ D_{\sigma^{+}_i}[\varrho] \right)$ \cite{LandiEtAl22},
%	\label{eq:Local-lind}
% \end{align}
 with rates $  \gamma_i^{\pm} = 2  W(\pm2h_i)$. For pure ohmic baths there is no lamb shift contribution to the Hamiltonian since $W(E)$ is real. 
 %Note that for the XXZ chain in absence of an external field, $h_i=0$, exact steady state solutions have been found for the local Lindblad equation \cite{Prosen11,Prosen11_2}.
The XXZ chain with local Lindblad equation has attracted a lot of attention because it can have exactly solvable steady states  \cite{Prosen11,Prosen11_2}.
 
   \begin{figure}
\centering
\includegraphics[scale=0.8]{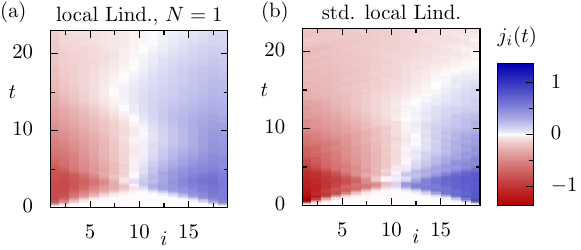}
\caption{Site-resolved magnetization current $j_i(t)$ (a)~for the local Linblad with order $N=1$ and (b)~for the traditional local Lindblad equation. Results obtained from quantum trajectory unraveling for the parameters of Fig.~\ref{fig:locallind}(d). }
\label{fig:qtra}
\end{figure}
 
As we observe in Fig.~\ref{fig:error-rf}(a), the populations of the many-body eigenstates are only poorly described by the standard local Lindblad equation. This poor performance is confirmed by Fig.~\ref{fig:error-rf}(c),(d) and Fig.~\ref{fig:locallind}(a) where we plot the measure $d$ for the distance of its steady state to %of the local Lindblad and 
the exact Redfield steady state. %and again for the approximate Redfield and Lindblad master equations from Fig.~\ref{fig:error-rf}(c),(d) but for fixed $\varepsilon_0 = 0.6J$. 
 Indeed, it is expected since one way to derive the standard local Lindblad eq. microscopically \cite{LandiEtAl22} is  in the limit  $J\to  0$. Nevertheless, such local jumps have attracted considerable attention \cite{Znidaric2010,Zanoci2023}, also in regimes where there is no microscopic justification \cite{Prosen11,Prosen11_2,LandiEtAl22}, as well as in the context of collisional models \cite{CattaneoEtAl21,CiccarelloEtAl22}. %, simply because for many-body systems only a limited amount of approaches were available. 
Our local Lindblad eq.\ is an alternative that has similar properties, but can overcome some of its flaws: As we show in Fig.~\ref{fig:locallind}(b), it predicts a  steady-state magnetization profile that is much closer to the actual thermal result predicted by the Redfield equation \cite{SM}. In Fig.~\ref{fig:locallind}(c) we additionally show the site-resolved magnetization current \cite{LandiEtAl22},
$ %\begin{align}
j_i(t)= -i \langle [H_S^{i,i+1}, \sigma_i^z]\rangle
$ %\end{align}
with $H_S^{i,i+1}$ denoting the nearest-neighbor interaction term in Eq.~\eqref{eq:H-xxz} at site $i$.
We observe that even though the system is connected to two heat baths with identical temperature $T_l=T_r =T$, the standard local Lindblad master equation, due to the tilted external field $h_i$, predicts a finite steady-state current through the system which is unphysical. The Redfield, local Redfield and our local Lindblad eq.\ do not suffer from such unphysical behavior \footnote{our local Lindblad eq.~only slightly violates the correct behavior at the bouandaries  \cite{SM}}. %, however the violation is less severe than what is predicted by the traditional local Lindblad approach. %Additionally in the SM \cite{SM} we show that the approximate Lindblad eq. shows thermalization of the many-body eigenstates and 
Hence, our local Lindblad eq.~fulfills all criteria for a consistent Markovian master equation that were formulated in Ref.~\cite{TupkaryEtAl23}:  It obeys 1) complete positivity, 2) local conservation laws and 3) shows thermalization. In  the SM \cite{SM}, we also show that our local master equations describe nonequilibrium setups (where $T_l\neq T_r$) with similar high accuracy.
% Note t

 \emph{Quantum trajectories for local Lindblad form}---We combine our new and the traditional local Lindblad eq.\ with established quantum trajectory methods \cite{KMolmer1993,GardinerEtAl92,ADaley2014,LandiEtAl22}, where we implement the algorithm of Ref.~\cite{KMolmer1993}. Since both forms yield jump operators that have a sparse representation, they allow us to treat relatively large systems. In Fig.~\ref{fig:locallind}(d) and \ref{fig:qtra} we demonstrate this by increasing the length of the spin chain to $L=19$, where the Hilbert space dimension is $2^L= 524288$. With global master equations such system sizes are numerically out of reach, %since both computational time and memory for diagonalization of $H_\mathrm{S}$, storage of the dense jump operator $u$ and propagation with the dense operator $u$ 
 and typically limit one to Hilbert space dimensions of $\sim 10^4$  on current hardware \cite{BeckerEtAl22,BeckerEtAl23}. We observe again that the magnetization profile $\langle \sigma_i^{z} \rangle$ in Fig.~\ref{fig:locallind}(d) that is predicted by the traditional local Lindblad approach deviates strongly from the more accurate novel local Lindblad dynamics.  Also the traditional local Lindblad equation overestimates the magnetization currents that we plot in Fig.~\ref{fig:qtra} in the initial relaxation dynamics. 
 %At later times, there are current oscillations in the middle of the chain that are not captured by the standard local Lindblad eq.~and its steady state again suffers more strongly from unphysical finite currents. 
 %that occur even though the system should reach equilibrium. 
%Our method might also serve as an alternative to exact diagonalization via Lanczos, or to quantum Monte Carlo methods in order to calculate thermal states. 
%This is true especially at intermediate temperatures $T$ where the number of many-body eigenstates that have significant contribution to the canonical Gibbs state scales generally with the Hilbert space dimension.
%Note that our approach is conceptually much simpler than combining the Redfield eq.~with matrix product operator methods as proposed in Ref.~\cite{XuEtAl19}, while %not requiring additional approximations and being able to simulate similar system sizes. 
By combining our approach with matrix product state methods for quantum trajectories, we expect that even larger system sizes are attainable.

\emph{Summary.}---Even though the Redfield master equation is typically considered  a global master equation, we emphasize that for coupling to local baths with short bath-correlation time, the jump operator will remain localized in space. We provide a Taylor-like expansion of the Redfield jump operator in terms of local operators, and show convergence for high and intermediate bath temperatures. This additionally has the advantage that a diagonalization of the full Hamiltonian is not required.
However, there might exist other expansions \cite{Shiraishi24} with better convergence especially in the case of bath spectral densities with a low cutoff. We show that our method can be further approximated to a Lindblad form with local jump operators, outperforming the traditonal local Lindblad ansatz. We combine our results with quantum trajectories to simulate the Redfield dynamics for Hilbert space dimensions that were previously challenging. Our method can be augmented with mean-field- or matrix-product methods to solve the Redfield equation for extended many-body systems.

%TODO: Show that leading order leads to Brownian motion

%TODO: DO MEAN FIELD for Redfield, compare against exact profile, current....
\begin{acknowledgments}
We acknowledge discussions with Igor Lesanovsky, Tobias Becker, Juzar Thingna, Francesco Petiziol, Gabriel Landi, Michiel Wouters, and Zala Lenar\v{c}i\v{c}. We are grateful  for comments on the manuscript by André Eckardt and  Archak Purkayastha.
This work was supported by the Deutsche Forschungsgemeinschaft (DFG, German Research Foundation) via the Reasearch Unit FOR 5688 (Project No.\ 521530974).
\end{acknowledgments}

\bibliography{mybib}

\end{document}

\title{Supplemental Material for ``Global becomes local: Efficient many-body dynamics for global master equations''}
\author{Alexander~Schnell} 
\email[Electronic address: ]{schnell@tu-berlin.de}
\affiliation{Technische Universit{\"a}t Berlin, Institut f{\"u}r Physik und Astronomie, 10623 Berlin, Germany}

\date{\today}

\maketitle
\subsection{Redfield equation for general (non-product) bath interaction Hamiltonian}

Using a Schmidt decomposition of a general given interaction Hamiltonian $H_\SB$, we find \cite{BreuerPetruccione,Schnell2019PHD,Thingna2013PHD}
\begin{align}
 H_\SB = \sum_{\alpha} v_\alpha \otimes B_\alpha.
\end{align}
If the bath is coupled locally around site $i_0$, all of the system coupling operators  $v_\alpha$ will be localized   around site $i_0$.
%where we require that the  system-  and bath coupling operators are Hermitian, $s_\alpha=s_\alpha^\dagger$, $B_\alpha= B_\alpha^\dagger$. 
The Redfield master equation then reads
\begin{align}
\partial_t {\varrho}(t) =  - {i} \left[H_\mathrm{S}, \varrho(t)\right] +  \sum_{\alpha}  \biggl( \left[  u_{\alpha} {\varrho}(t), {v}_\alpha\right]  + \mathrm{h.c.} \biggr),
\label{eq:BornMarkov-noRWA}
\end{align}
with Redfield jump operators
\begin{align}
u_{\alpha} =  \sum_{ \beta}  \int_0^\infty \! \mathrm{d}\tau  \tilde{v}_\beta(-\tau) C_{\alpha\beta}(\tau),
\label{eq:u-general}
\end{align}
and bath correlation functions
\begin{align}
C_{\alpha\beta}(t) = \left\langle \tilde{B}_\alpha(t)\tilde{B}_\beta(0) \right\rangle_\mathrm{B}.
\end{align}
Starting from the sum in Eq.~\eqref{eq:u-general}, we can apply our arguments of the main text to each of the terms in the sum individually to derive a local Redfield master equation.

\subsection{Convergence criterion for ad-hoc series expansion}
\label{sec:taylor-conv}

We would like to formulate criteria under which the ad-hoc expansion of the Redfield jump operator $u$, Eq.~(3) in the main text, converges.
Here we show that strict convergence is always guaranteed if all transition energies $E_k-E_q$ with eigenenergies $E_k$ of the system Hamiltonian $H_\mathrm{S}$ lie in the convergence radius of the Taylor series of $W(E)$ around $E=\varepsilon_0$. This is always true for ohmic baths at temperatures $T$ that are higher than the  transition energies that are allowed by system coupling operator $v$.
For an interacting many-body system, the full energy spectrum is potentially unknown and broadening with system size. However, when the coupling operator $v$ is a local operator it suffices to require Taylor convergence for  transition energies which  are bounded by the energy cost for local operations.

\begin{figure}
\centering
\includegraphics[scale=0.8]{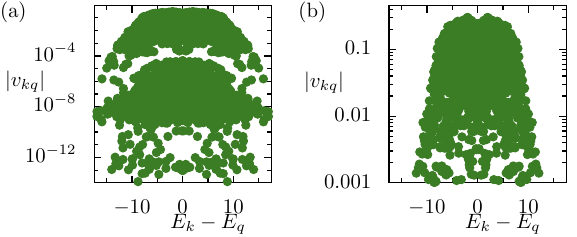}
\caption{(a)~Magnitude of the coupling matrix elements $v_{kq} = \bra{k}v\ket{q}$ against energy difference $E_k - E_q$ (in units of $h$) for the left coupling operator $v=v_l$ of the spin chain in Fig.~2(a) of the main text for $L=6$. (b) Zoomed in version of (a).} 
\label{fig:couplmatelm}
\end{figure}

\begin{figure}
\centering
\includegraphics[scale=0.8]{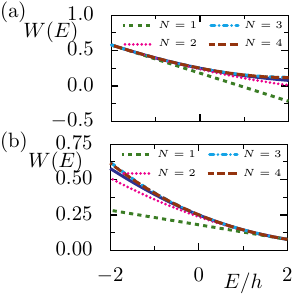}
\caption{Convergence of the Taylor series of  $W(E)$ around (a) $\varepsilon_0 = -2h$, (b) $\varepsilon_0 =2h$ for of a pure ohmic bath at temperature $T=h$ (solid blue line). } 
\label{fig:conv-W_E}
\end{figure}

This can be seen from introducing the energy eigenstates $H_\mathrm{S}\ket{k} = E_k \ket{k}$ in
\begin{align}
u &= \int_0^\infty \mathrm{d}\tau \tilde{v}(-\tau) C(\tau)\\
&=\sum_{kq} \int_0^\infty \mathrm{d}\tau \ket{k}\bra{k}\tilde{v}(-\tau) \ket{q}\bra{q} C(\tau).
\end{align}
We then use $\bra{k}\tilde{v}(-\tau) \ket{q}= \exp(-i(E_k-E_q)\tau) \bra{k}{v}\ket{q}$ and define the eigenspace-projected 
coupling operator $\hat{v}_{kq} = \ket{k}\bra{k}{v}\ket{q}\bra{q}$ and the Fourier-Laplace transform of the bath correlation function
\begin{align}
W(E) &= \int_0^\infty \mathrm{d}\tau \mathrm{e}^{-iE\tau} C(\tau).
\end{align}
This yields the from
\begin{align}
u 
&=\sum_{kq}W(E_k-E_q) \hat{v}_{kq} .
\label{eq:u-op-basis}
\end{align}
Assuming that $W(E)$ is analytical and all transition energies $E_k-E_q$ are in the convergence radius of the Taylor series of $W(E)$ around $E=\varepsilon_0$ we can rewrite exactly (no approximation)
\begin{align}
W(E) &= \sum_{n=0}^{\infty} \frac{W^{(n)}(\varepsilon_0)}{n!} (E-\varepsilon_0)^n.
\end{align}
Plugging this into Eq.~\eqref{eq:u-op-basis}, we obtain 
\begin{align}
u 
&=\sum_{n=0}^{\infty}  \sum_{kq} \frac{W^{(n)}(\varepsilon_0)}{n!} (E_k-E_q-\varepsilon_0)^n \hat{v}_{kq} .
\end{align}
After using that $(E_k-E_q-\varepsilon_0) \hat{v}_{kq} = ([H_\mathrm{S}, \cdot ] -\varepsilon_0 \cdot )[\hat{v}_{kq}]$, 
we can remove the energy basis and find that
\begin{align}
u 
&=\sum_{n=0}^{\infty} \frac{W^{(n)}(\varepsilon_0)}{n!} ([H_\mathrm{S}, \cdot ] -\varepsilon_0 \cdot )^n [v].
\label{eq:adh-locred}
\end{align}
Hence we have shown convergence of the series in Eq.~(4) for the case that all transition energies are in the convergence radius of the scalar Taylor series around $\varepsilon_0$.

\begin{figure}
\centering
\includegraphics[scale=0.8]{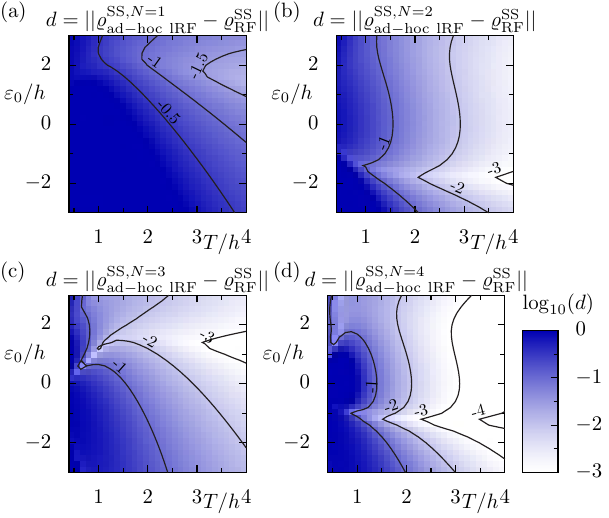}
\caption{Trace distance $d$ of the steady state of the exact Redfield and the ad-hoc local Redfield master equation truncated at order (a)~$N=1$, (b)~$N=2$, (c)~$N=3$, (d)~$N=4$, as function of temperature $T$ and expansion energy $\varepsilon_0$. Parameters as in Fig.\ 2(a) of the main text.}
\label{fig:supp-rf-ord}
\end{figure}

By looking at Eq.~\eqref{eq:u-op-basis}, we observe that this criterion can be relaxed for all transitions where  $\hat{v}_{kq}\approx 0$. In the case of $v$ being a local operator, this means that only a subset of all energy transitions occur with a significant amplitude $\vert {v}_{kq}\vert$. 
In Fig.~\ref{fig:couplmatelm} we plot the matrix elements of the  left local coupling operator $v=v_l$ against the energy differences in the case of the spin chain of the main text. We observe that, although the total energy differences $E_k-E_q$ approximately all lie in the interval $[-17h, 17h]$, the most relevant ones with $\vert v_{kq}\vert \geq 10^{-3}$ are dominantly confined to $[-10h, 10h]$. It therefore suffices to require convergence of the Taylor series in that interval.

Note that for ohmic baths we have
\begin{align}
W(E) = \frac{\gamma E}{\exp(E/T)-1},
\end{align}
whose graph and Taylor series we plot in Fig.~\ref{fig:conv-W_E}. Its poles lie on the imaginary axis and are given by the Matsubara frequencies, $E_n = i \nu_l =i 2\pi  T l$, $l\in \mathbb{Z}\backslash\{0\}$. As a result, the convergence radius of a Taylor series around $\varepsilon_0=0$ is $2\pi T$. Hence as long as $T\gg \vert E\vert/2\pi$ for all relevant transition energies $E$, we expect that the ad-hoc local Redfield equation converges. Therefore here, as long as $T\gg 10h/(2\pi)  \approx 1.6 h$, we expect convergence of the ad-hoc local Redfield equation for $\varepsilon_0=0$. This is confirmed by the convergence of the steady-state error of the ad-hoc local Redfield equation in Fig.~\ref{fig:supp-rf-ord}, as we discuss in more detail in Sec.~\ref{sec:app-error-converg}.

\subsection{Local Redfield equation: Application to Bose-Hubbard chain}
To generalize our results also to other quantum systems, we assume a generic quantum system, whose Hamiltonian can almost always be understood as the sum of purely onsite terms $H_\sys^{\mathrm{loc}, i}$ that only act on site $i$, and the residual interaction- or tunneling terms, i.e.
\begin{align}
H_\sys =  \sum_{i=1}^{L} H_\sys^{\mathrm{loc}, i} + H_\sys^{\mathrm{resid}}.
\label{eq:H-gener}
 \end{align}
%In case of the XXZ Hamiltonian of the main text, e.g. this decompositon reads
%  \begin{align}
%H_\sys ={ -J \sum_{i=1}^{L-1}\left( \sigma^x_{i} \sigma^x_{i+1} +  \sigma^y_{i} \sigma^y_{i+1} + \Delta  \sigma^z_{i} \sigma^z_{i+1}\right)} + \sum_{i=1}^{L} \underbrace{h_i \sigma^z_{i}}_{H_\sys^{\mathrm{loc}, i}}.
%\label{eq:H-xxz-supp}
% \end{align}
 As another example, here we address a Bose-Hubbard Hamiltonian with external trapping potential $V_i$, 
   \begin{align}
H_\sys ={ -J \sum_{i=1}^{L-1}\left( \hat a^\dagger_{i+1} \hat a_i+\hat a^\dagger_{i} \hat a_{i+1}\right)} + \sum_{i=1}^{L}  \underbrace{\left[ \frac{U}{2}  \hat  n_i ( \hat  n_i+1)+V_i  \hat n_i \right]}_{H_\sys^{\mathrm{loc}, i}}.
 \end{align}
%Even though exact diagonalisation is generally a hard task for interacting many-body systems, 
As mentioned in the main text,
this site-local Hamiltonian can typically be diagonalized straightforwardly,
\begin{align}
H_\sys^{\mathrm{loc}, i} \ket{k^{i}}= E^{\mathrm{loc},i}_k \ket{k^{i}},
 \end{align}
by introducing the on-site Fock basis
\begin{align}
\hat n_i \ket{\lbrace n_i \rbrace }= n_i \ket{\lbrace n_i \rbrace },
 \end{align}
 with $n_i \in \mathbb{N}_0$. This gives 
 \begin{align}
H_\sys^{\mathrm{loc}, i}  \ket{\lbrace n_i \rbrace }&= E^{\mathrm{loc},i}_{n_i}  \ket{\lbrace n_i \rbrace }\\
&= \left[ \frac{U}{2}    n_i (  n_i+1)+V_i   n_i  \right] \ket{\lbrace n_i \rbrace }.
 \end{align}
 
 On site $i=1$, we couple the system to a bosonic particle reservoir with bath Hamiltonian 
 \begin{align}
H_\mathrm{B} =  \sum_{i} E^\mathrm{B}_i \hat b^\dagger_i \hat b_i,
 \end{align}
 using the system-bath coupling Hamiltonian
 \begin{align}
H_\mathrm{SB} =  \sum_{i} c_i \left(\hat  a_1 \hat b^\dagger_i + \hat a^\dagger_1 \hat b_i \right) = \sum_{i=1}^2 v_i \otimes B_i.
 \end{align}
 Here we have defined $v_1 = \hat a_1$, $B_1 = \sum_i c_i \hat b^\dagger_i$ as well as  $v_2 = \hat a^\dagger_1$, $B_2 = \sum_i c_i \hat b_i$.
 
 \begin{figure}
\centering
\includegraphics[scale=0.77]{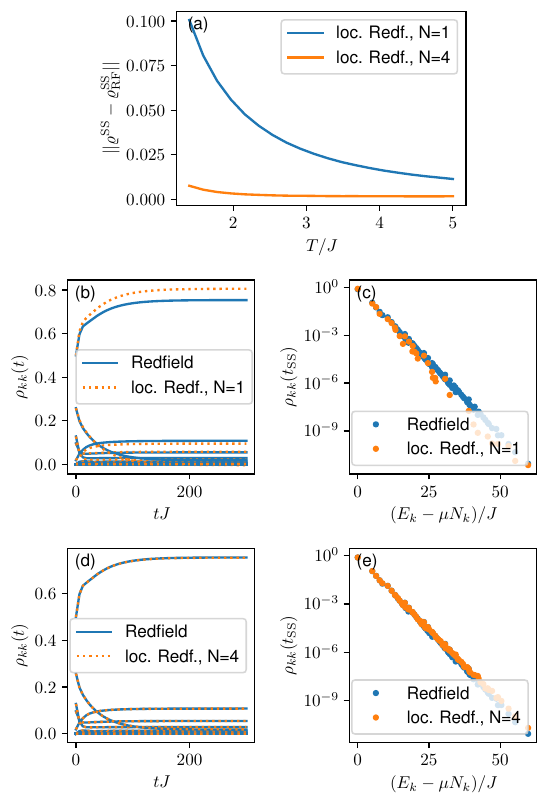}
\caption{(a) Steady-state trace error of the local Redfield master eq.~for a Bose-Hubbard chain of length $L=4$ with $U=0.5J$, $V_i = 0.35 (i-L/2)^2 J$, coupled to thermal particle reservoirs with $\mu=-2.5 T$ and $\gamma =0.25J$ for orders $N=1,4$. We cut off the bosonic on-site populations by $n_i \leq 2$.
(b)~Dynamics and (c) steady-state values of the populations of the many-body eigenstates $k$ for order $N=1$ at temperature $T=2J$. (d) and (e): Same as in  (b) and (c) but for order $N=4$.} 
\label{fig:BoseHubbard}
\end{figure}
 
 This choice yields the Redfield master equation
 \begin{align}
\partial_t {\varrho}(t) =  - {i} \left[H_\mathrm{S}, \varrho(t)\right] &+   \biggl( \left[  u_{1} {\varrho}(t), \hat a_1\right]  + \mathrm{h.c.} \biggr)\\
&+   \biggl( \left[  u_{2} {\varrho}(t), \hat a^\dagger_1\right]  + \mathrm{h.c.} \biggr),
\end{align}
where
 \begin{align}
u_{1} & = \sum_{\beta=1}^2  \int_0^\infty \! \mathrm{d}\tau  \tilde{v}_\beta(-\tau) C_{1\beta}(\tau)\\
	& =   \int_0^\infty \! \mathrm{d}\tau\tilde{a}^\dagger_1(-\tau)  \sum_{i} c_i^2 \langle b^\dagger_i(\tau) b_i\rangle, \\
u_{2} & =   \int_0^\infty \! \mathrm{d}\tau\tilde{a}_1(-\tau)  \sum_{i} c_i^2 \langle b_i(\tau) b^\dagger_i\rangle.
\end{align}
In analogy to the XXZ chain in the main text, we introducing the functions
 \begin{align}
W_1(E) &=  \int_0^\infty \! \mathrm{d}\tau e^{-iE\tau} \sum_{i} c_i^2 \langle b^\dagger_i(\tau) b_i\rangle\\
& = J(E) n_\mathrm{B}(E),\\
W_2(E) &=  \int_0^\infty \! \mathrm{d}\tau e^{-iE\tau} \sum_{i} c_i^2 \langle b_i(\tau) b^\dagger_i\rangle\\
& = J(-E)(1+ n_\mathrm{B}(-E)),\\
 \end{align}
 where in the second steps, neglected the imaginary parts of $W_1(E)$, $W_2(E)$ and 
  we have introduced the spectral density,  
 \begin{align}
J(E) &= \sum_{i} c_i^2 \delta(E- E^\mathrm{B}_i) = \gamma
 \end{align}
  which for the particle reservoir of the Bose-Hubbard system, we assume a coupling to a flat spectral density. Also it enters the Bose-Einstein distribution 
   \begin{align}
n_\mathrm{B}(E) &= \frac{1}{e^{(E-\mu)/T}-1}
 \end{align}
 with bath temperature $T$ and -chemical potential $\mu$.
 This finally yields the exact Redfield operators
  \begin{align}
u_{1} & = \sum_{kq}\ket{k}\bra{k} a^\dagger_1 \ket{q}\bra{q} W_1(E_k-E_q) , \\
u_{2} & =  \sum_{kq}\ket{k}\bra{k} a_1 \ket{q}\bra{q} W_2(E_k-E_q).
\end{align}
Here we have introduced the exact many-body eigenstates $H_\mathrm{S} \ket{k} = E_k \ket{k}$.

From this form, we have the local Redfield jump operators,
  \begin{align}
u_{1} & \approx \sum_{n=0}^N \sum_{n_1} \frac{W^{(n)}_1(\varepsilon^+_{n_1})}{n!} \left([H_\mathrm{S}, \cdot] -\varepsilon^+_{n_1} \cdot \right)^n [(a^\dagger_1)_{n_1+1,n_1}]  , \\
u_{2} &  \approx \sum_{n=0}^N \sum_{n_1} \frac{W^{(n)}_2(\varepsilon^-_{n_1})}{n!} \left([H_\mathrm{S}, \cdot] -\varepsilon^-_{n_1} \cdot \right)^n [(a_1)_{n_1-1,n_1}] ,
\end{align}
with $(a^\dagger_1)_{n_1+1,n_1} = \ket{n_1+1}\bra{n_1+1} a^\dagger_1 \ket{n_1}\bra{n_1}$ and  $(a_1)_{n_1-1,n_1} = \ket{n_1-1}\bra{n_1-1} a_1 \ket{n_1}\bra{n_1}$. Here enter the local transition energies
  \begin{align}
\varepsilon^+_{n_1} &= E^{\mathrm{loc},1}_{n_1+1}  - E^{\mathrm{loc},1}_{n_1} = U n_1 + V_1 , \\
\varepsilon^-_{n_1} &= E^{\mathrm{loc},1}_{n_1-1}  - E^{\mathrm{loc},1}_{n_1} = -U (n_1-1) - V_1.
\end{align}
In Fig.~\ref{fig:BoseHubbard}(a), we plot  the  steady-state error as function of temperature $T$ for orders $N=1$ and $N=4$ for the Bose-Hubbard chain coupled to the particle reservoir and we observe similar good agreement as in the case of the XXZ chain in the main text.
Note that in order to have roughly the same particle number at all temperatures, we scale the chemical potential with temperature.
In Fig.~\ref{fig:BoseHubbard}(b)-(e),  we additionally show the populations of the many-body eigenstates for the same parameters as  Fig.~\ref{fig:BoseHubbard}(a) at $T=2J$ during the dynamics and in  the steady state.

\subsection{Bath-correlation time for ohmic bath}
In Fig.~3(a) of the main text, we plot the error as a function of relative bath-correlation time  $\tau_\mathrm{B}$.
To that end, we want to define the bath correlation time for a pure ohmic bath. In order to have convergent integrals for the bath-correlation function, we introduce a Drude cutoff energy $E_\mathrm{D}$, giving the spectral density
\begin{align}
	J_\mathrm{D}(E) = \frac{\gamma E}{1+E^2/E_\mathrm{D}^2},
\end{align}
where in the end we will take the limit $E_\mathrm{D}\to \infty$. 
For a phonon bath as in the main text, the bath-correlation function reads as \cite{Thingna2013PHD}
\begin{align}
	C(t) = \int_0^\infty \frac{\mathrm{d}E}{\pi} J_\mathrm{D}(E) \left[\coth\left(\frac{E}{2T}\right) \cos(E t)-i \sin(E t)\right].
\end{align}
For the Drude-Lorentz  bath, this gives \cite{Thingna2013PHD}
\begin{align}
	\begin{split}
	C(t) =&\frac{\gamma}{2} E_\mathrm{D}^2 \left[\cot\left(\frac{E_\mathrm{D}}{2T}\right)-i \right] e^{-E_\mathrm{D}t}\\
	&-{2\gamma T} \sum_{l=1}^{\infty} \frac{\nu_l  e^{-\nu_l t}}{1-(\nu_l/E_\mathrm{D})^2}. \qquad (t>0)
	\end{split}
\end{align}
Here we have introduced the Matsubara frequencies  $\nu_l = 2\pi T l$. Note that the bath-correlation function $C(t)$ is the sum of multiple exponential decays. Naively, we define as the bath-correlation time $\tau_\mathrm{B}$ the longest time scale of decay of these exponentials (which is sensible at high  temperatures $T$). This finally yields $\tau_\mathrm{B}=(2\pi T)^{-1}$.

\begin{figure}
\centering
\includegraphics[scale=0.8]{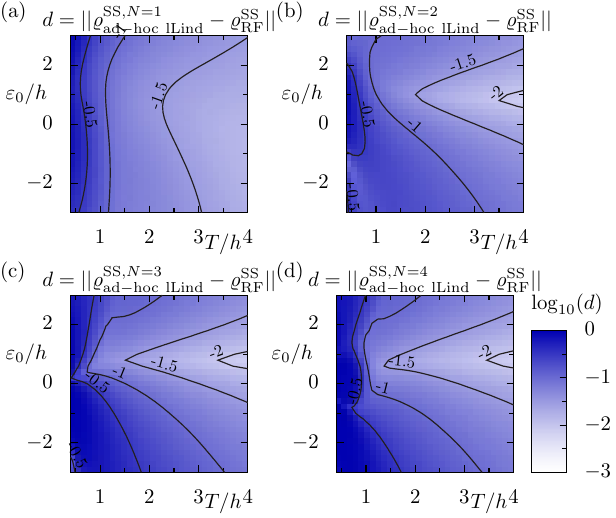}
\caption{Trace distance $d$ of the steady state of the exact Redfield and the ad-hoc local Lindblad master equation truncated at order (a)~$N=1$, (b)~$N=2$, (c)~$N=3$, (d)~$N=4$. Parameters as in Fig.~2(a) of the main text.}
\label{fig:supp-lind-ord}
\end{figure}

 \begin{figure}
\centering
\includegraphics[scale=0.8]{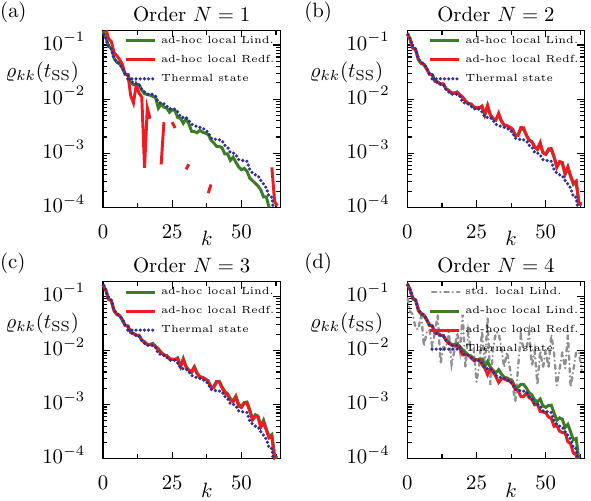}
\caption{Steady-state populations $\varrho_{kk}(t_\mathrm{SS})$ of the many-body eigenstates $k$ for the ad-hoc local Redfield- and Lindblad equations for $\varepsilon_0=0.6h$ against the exact thermal state given by the canonical ensemble $\varrho_\mathrm{th} = \exp(-H_\mathrm{S}/T)/Z$. In (a) some of the populations for the ad-hoc local Redfield are not drawn because they are negative. In (d) we also plot the steady state of the local Lindblad eq.~(which is independent of order $N$). Parameters of Fig.~2(b) of the main text.} 
\label{fig:thermaliz}
\end{figure}
 
 \begin{figure}
\centering
\includegraphics[scale=0.8]{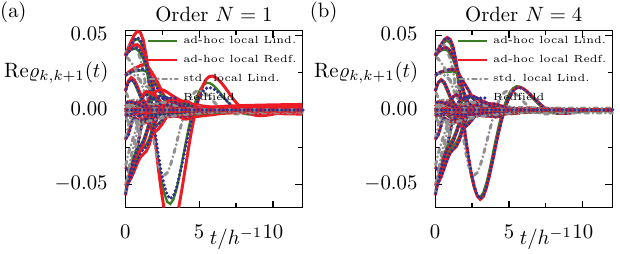}
\caption{Dynamics of the coherences between neighboring many-body eigenstates $k$ for the approximate master equations for $\varepsilon_0=0.6h$ and the parameters of Fig.~2(a) of the main text.} 
\label{fig:coher}
\end{figure}

\subsection{Rewriting Redfield master equation in pseudo-Lindblad form}

We follow Ref.~\cite{BeckerEtAl21} and rewrite the Redfield master eq.~(1) in the main text as
\begin{align}
  \partial_t \varrho = -i \left[H, \varrho \right] &+ [u \varrho, v] + [v, \varrho u^\dagger]\\
  \begin{split}
  = -i \left[H, \varrho \right] &+  u\varrho v - vu \varrho -\tfrac{1}{2} \varrho vu  +\tfrac{1}{2} \varrho vu\\
  &+ v \varrho u^\dagger - \varrho u^\dagger v  -\tfrac{1}{2} u^\dagger v \varrho  +\tfrac{1}{2} u^\dagger v \varrho 
  \end{split}\\
    \begin{split}
  = -i [H- \tfrac{i}{2}&(vu-u^\dagger v), \varrho] \\  %+ [u \varrho(t), v] + [v, \varrho(t) u^\dagger]
 & + \left(\!\begin{array}{cc}u & v \end{array}\!\right)\! \left( \begin{array}{cc}0 & \varrho \\
 \varrho & 0 \end{array}\right) \! \left(\!\begin{array}{cc}u^\dagger \\ v \end{array}\!\right) \\
 &-\tfrac{1}{2} \left\lbrace \left(\!\begin{array}{cc}u^\dagger & v \end{array}\!\right) \! \left( \begin{array}{cc}0 & 1 \\
1 & 0 \end{array}\right) \!  \left(\!\begin{array}{cc}u \\ v \end{array}\!\right),  \varrho  \right\rbrace.
\label{eq:redf-rewr}
   \end{split}
\end{align}
We note that for any $\eta \in \mathbb{C}$ we have
\begin{align}
\left( \begin{array}{cc}0 & 1 \\
1 & 0 \end{array}\right) = \Omega \left( \begin{array}{cc} \eta^{-1} & (\eta^*)^{-1}  \\
\eta & -\eta^* \end{array}\right)  \left( \begin{array}{cc}1 & 0 \\
0 & -1 \end{array}\right) \left( \begin{array}{cc} (\eta^*)^{-1} & \eta^* \\
\eta^{-1}  & -\eta \end{array}\right) 
\end{align}
with $\eta^*$ denoting the complex conjugate and normalization factor $\Omega=\vert \eta\vert^2 / [(\eta^*)^2 + \eta^2]$.

By plugging this into Eq.~\eqref{eq:redf-rewr}, we can rewrite the Redfield equation into pseudo-Lindblad form \cite{BeckerEtAl21}
\begin{align}
	\begin{split}
	  \partial_t \varrho  = -i [H- \tfrac{i}{2}&(vu-u^\dagger v), \varrho] + L_+ \varrho L_+^\dagger - \tfrac{1}{2} \left \lbrace L_+^\dagger L_+, \varrho \right \rbrace \\
	 & - \left( L_- \varrho L_-^\dagger - \tfrac{1}{2} \left \lbrace L_-^\dagger L_-, \varrho \right \rbrace\right)
	 \end{split}
\end{align}
where the term \emph{pseudo}-Lindblad refers the presence of a negative prefactor (i.e.~a negative jump rate) for the second term in the dissipator.
The jump operators read
\begin{align}
	L_+ &= \tfrac{\vert \eta \vert}{\sqrt{\eta^2+(\eta^*)^2}}\left[ \eta^{-1} u + \eta v \right],\\
	L_- &= \tfrac{\vert \eta \vert}{\sqrt{\eta^2+(\eta^*)^2}}\left[ (\eta^*)^{-1} u - \eta^* v \right].
	\label{eq:Lminus}
\end{align}
By rewriting $L_-$ we find
\begin{align}
	L_- &= \tfrac{1}{\vert \eta \vert} \tfrac{1}{\sqrt{1+(\eta^*)^2/\eta^2}}  \left[ u - (\eta^*)^2 v \right].
	\label{eq:Lminus_new}
\end{align}
Since according to Eq.~(3) in the main text, on leading order $N=0$,  $u^{(0)}=W(\varepsilon_0) v$, we see that the  leading order contribution to $L_-$ for the ad-hoc local Redfield eq.~can be eliminated by setting $(\eta^*)^2=W(\varepsilon_0)$.

This finally yields the jump operators
\begin{align}
	L_+ &= \tfrac{1}{\sqrt{\vert W(\varepsilon_0) \vert}} \tfrac{1}{\sqrt{1+W(\varepsilon_0)^*/W(\varepsilon_0)}}  \left[ u + W(\varepsilon_0)^* v \right],\\
	L_- &=\tfrac{1}{\sqrt{\vert W(\varepsilon_0) \vert}} \tfrac{1}{\sqrt{1+W(\varepsilon_0)/W(\varepsilon_0)^*}}  \left[ u - W(\varepsilon_0) v \right],
\end{align}
which simplify to the expressions in the main text for $W(\varepsilon_0)\in \mathbb{R}$. By neglecting the jump-operator $L_-$, we find the ad-hoc local Lindblad equation.

\subsection{Error of ad-hoc local master equations for different truncation orders}
\label{sec:app-error-converg}
In Fig.~\ref{fig:supp-rf-ord} we plot the steady state errors $d$ as defined in the main text for different orders of the cutoff $N=1,2,3,4$ of the ad-hoc local Redfield eq.~(3) in the main text for the spin chain with ohmic baths. As discussed in Sec.~\ref{sec:taylor-conv}, we observe fast convergence for intermediate and high temperatures $T \gtrsim 2h$ for a wide range of expansion energies $\varepsilon_0$. For low $T$, the expansion does not seem to converge, as is also indicated by the rapid increase of the error at around $T \approx 2h$ in Fig.~\ref{fig:supp-rf-ord}(d).

Similarly, in Fig.~\ref{fig:supp-lind-ord}, we plot the steady state errors $d$ for the ad-hoc local Lindblad equation
\begin{align}
\partial_t \varrho 
	= -i \left[ H_\mathrm{eff}, \varrho \right] + D_{L_+}[\varrho],
\end{align}
with effective Hamiltonian $H_\mathrm{eff} = H_\sys-\frac{i}{2}(vu-u^\dagger v)$ and  Lindblad superoperator $D_{L}[\varrho]=\left( L\varrho L^\dagger - \frac{1}{2} \lbrace L^\dagger L,\varrho \rbrace\right)$ with jump operator (since $W(E)$ is real)
\begin{align}
L_+ = \frac{1}{\sqrt{2W(\varepsilon_0)}} \left[ u + W(\varepsilon_0) v\right]
\end{align}
 where we expand the operator $u$ according to the ad-hoc expansion at different orders $N=1,2,3,4$.
By throwing away $L_-$ we make an error on order $(\Delta/T)^2$, so the ad-hoc local Lindblad form should be correct on order $N=1$ i.e.~$\Delta/T$ only. 
We observe that, first, surprisingly on this order $N=1$ the ad-hoc local Lindblad form performs much better than the ad-hoc Redfield master equation. This is due to the fact that the Lindblad dynamics confines the populations to values between $0$ and $1$, while in the Redfield dynamics they can take also large negative values.
Second, even though the Lindbladian approximation is consistent on order $N=1$ only, we observe that the accuracy can still be improved slightly by going to $N=2$ and $N=3$. Nevertheless going to $N=4$ and higher will not significantly improve the performance of the approximate Lindblad form anymore.

In Fig.~\ref{fig:thermaliz} we additionally plot the steady-state populations of the approximate master equations for different truncation orders $N$. We observe an excellent agreement with the thermal populations at each truncation order $N$, except for the ad-hoc local Redfield eq.~at $N=1$, where negative populations occur.
Interestingly, already for $N=1$ the ad-hoc local Lindblad equation shows an outstandingly good agreement with the exact thermal state.
% In a recent work \cite{TupkaryEtAl23} three criteria for a consistent Markovian master equation were formulated:  It obeys 1) complete positivity, 2) local conservation laws and 3) shows thermalization. While the standard Redfield eq.~violates criterion 1, the quantum optical ME violates criterion 2, and traditional local Lindblad eqs.~violate criterion 3.
%The ad-hoc local Lindblad master eq.~that we propose here fulfills all three criteria, however at the expense of possible approximation errors due to the truncation. Interestingly in Ref.~\cite{TupkaryEtAl23}, to find such an equation, the authors rely on a numerical optimization scheme. They find  that such an equation only exists if the corresponding Lindblad jump operators are acting over at least two sites. Consequentially, our work can provide an analytical derivation of the master equation that was targeted in that work numerically.

Finally, in Fig.~\ref{fig:coher} we plot the dynamics of some of the coherences of the many-body density matrix and observe good agreement with the exact result. Moreover, the ad-hoc local Lindblad master equation again performs well at $N=1$ already.

 \begin{figure}
\centering
\includegraphics[scale=0.8]{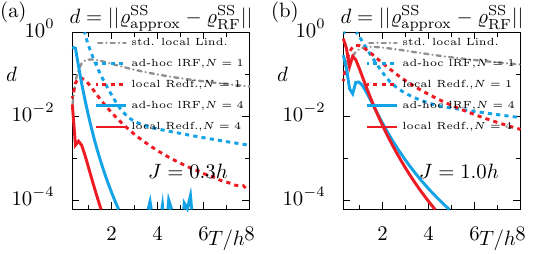}
\caption{Comparison of the steady state distance $d$ for the ad-hoc- and the local Redfield equation for (a) weak interaction $J=0.3h$ and (b) stronger interactions $J=h$ and $N=1$ (dashed), $N=4$ (solid). For the ad-hoc local Redfield eq.~we choose $\varepsilon_0=2J$.} 
\label{fig:dist-comp}
\end{figure}
 
  \begin{figure}
\centering
\includegraphics[scale=0.8]{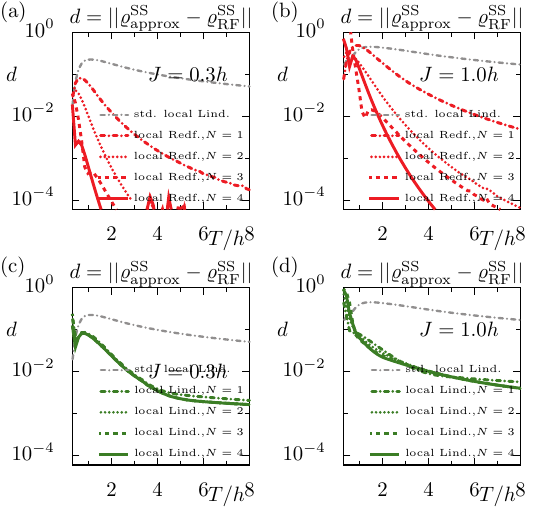}
\caption{Steady state distance $d$ for  (a),(b) the local Redfield equation and (c),(d) the local Lindblad equation for different orders of approximation $N=1,2,3,4$.  (a),(c) weak interaction $J=0.3h$ and (b),(d) stronger interactions $J=h$.} 
\label{fig:dist-multiple}
\end{figure}

\subsection{Local Redfield equation: Other orders}

 \begin{figure}
\centering
\includegraphics[scale=0.8]{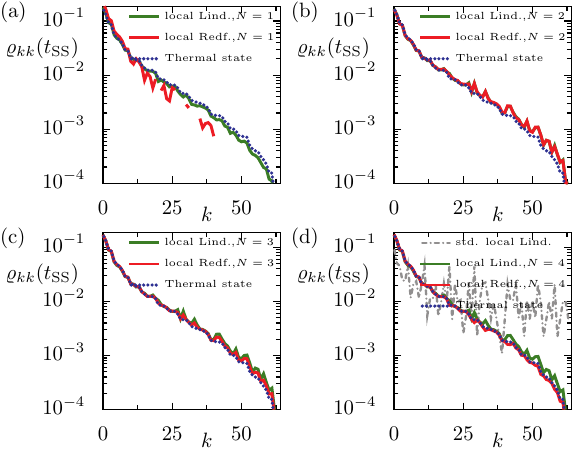}
\caption{Steady-state populations $\varrho_{kk}(t_\mathrm{SS})$ of the many-body eigenstates $k$ for the local Redfield- and Lindblad equations for orders $N=1$ to $N=4$ (a)-(d) against the exact thermal state given by the canonical ensemble $\varrho_\mathrm{th} = \exp(-H_\mathrm{S}/T)/Z$.}  
\label{fig:thermaliz-local}
\end{figure}
 
 \begin{figure}
\centering
\includegraphics[scale=0.8]{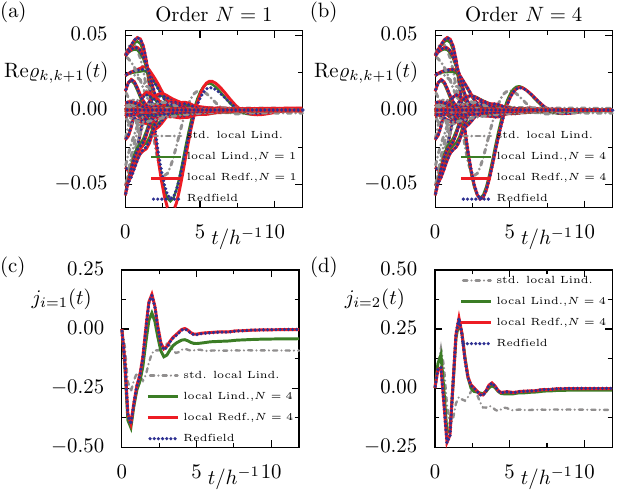}
\caption{(a),(b) Dynamics of the coherences between neighboring many-body eigenstates $k$ for the local master equations for the parameters of Fig.~2(a) of the main text. (c),(d) Local magnetization currents as defined in the main text for the different master equations at site (c) $i=1$ and (d) $i=2$.} 
\label{fig:coher-local}
\end{figure}

 We expect that, as long as $H_\sys^{\mathrm{resid}} \ll H_\sys$, the local Redfield equation, Eq.~(4) in the main text, converges faster than the ad-hoc local Redfield eq.~(3) in the main text.
This is confirmed by Fig.~\ref{fig:dist-comp} where we compare the steady-state error for the ad-hoc- and the local Redfield master equation at weak ($J=0.3h$) and strong ($J=h$) spin interaction. For weak $J$, the local Redfield equation outperforms the ad-hoc Redfield equation by orders of magnitude, cf. Fig.~\ref{fig:dist-comp}(a).

In Fig.~\ref{fig:dist-multiple}(a),(b) we additionally show the steady-state error also for other expansion orders $N=2,3$ of the local Redfield equation. We observe that initial convergence is very fast, so that for high and intermediate temperatures $T$, for practical purposes, it might suffice to resort to order $N=1$ or $N=2$ of the expansion.

\subsection{Details on local Lindblad equation}

In order to minimize the operator $L_-$, Eq.~\eqref{eq:Lminus}, for the local Redfield master eq.~for general quantum systems, we take inspiration from the recipe from Ref.~\cite{BeckerEtAl23} (which they call state-dependent optimization). They suggest the optimal choice for a pure state $\ket{\psi}$ during a quantum trajectory unraveling is
  \begin{align}
\eta_\mathrm{opt,BNE}^2 = \frac{\vert \vert u \ket{\psi} \vert \vert}{\vert \vert v \ket{\psi} \vert \vert},
 \end{align}
 which follows from minimizing the action of $L_-$ in Eq.~\eqref{eq:Lminus_new} on state $\ket{\psi}$. Here $\vert\vert \cdot \vert\vert$ corresponds to the 2-norm.

\begin{figure}
\centering
\includegraphics[scale=0.8]{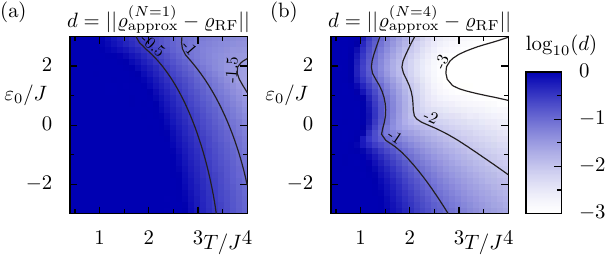}
\caption{Like in Fig.~\ref{fig:supp-rf-ord} but for Lorentz-Drude bath with $E_\mathrm{D}=30h$ (note that $J=h$).}
\label{fig:supp-rf-ord-drude-30}
\end{figure}

 \begin{figure}
\centering
\includegraphics[scale=0.8]{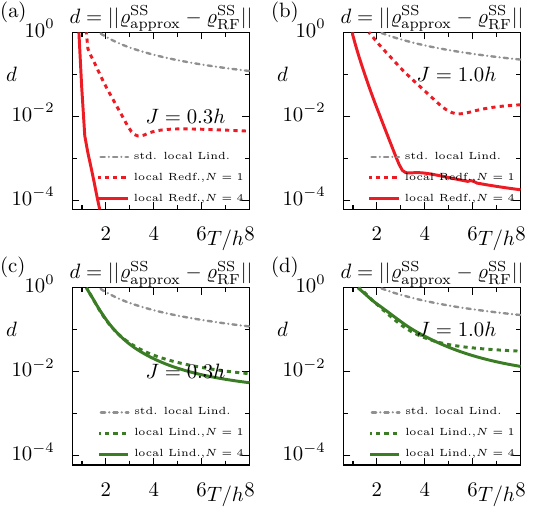}
\caption{Similar to Fig.~\ref{fig:dist-multiple} for  a Lorentz-Drude bath with $E_\mathrm{D}=30h$.} 
\label{fig:dist-multiple-drude-30}
\end{figure}
 
In our work, we aim to use a state-independent optimization. Nevertheless, we find the state-independent optimal parameter $\eta$ from Ref.~\cite{BeckerEtAl23} is not optimal.  We find that in our case it can be further optimized by considering the 
action of the leading order ($N=0$) of the local Redfield operator $u$ in $L_-$ on a local thermal steady state, giving
  \begin{align}
(\eta_\mathrm{opt}^*)^2 = \frac{\vert \vert \sum_{kq}  W(E^{\mathrm{loc},i_0}_k -E^{\mathrm{loc},i_0}_q)   \hat{v}^{i_0}_{kq} e^{-E^{\mathrm{loc},i_0}_q/T}\vert \vert_\mathrm{pseud}}{\vert \vert \sum_{kq}   \hat{v}^{i_0}_{kq} e^{-E^{\mathrm{loc},i_0}_q/T} \vert \vert_\mathrm{pseud}},
 \end{align}
 with a `pseudo' norm (maps to complex values),
  \begin{align}
\vert \vert A \vert \vert_\mathrm{pseud}= \sqrt{\sum_{ij} A_{ij}^2}.
 \end{align}
Since this optimal parameter only relies on the site-local eigenstates and -energies it can be easily calculated also for large many-body systems.

In Fig.~\ref{fig:dist-multiple}(c),(d) we show the steady-state error for this local Lindblad equation for expansion orders $N=1,2,3,4$ and weak- and strong interaction $J$. We observe that already at order $N=1$ the local Lindblad equation is essentially converged and cannot be improved by going to higher orders.

In Fig.~\ref{fig:thermaliz-local}  and Fig.~\ref{fig:coher-local}(a),(b) we again plot the steady-state populations and the dynamics of the coherences of the local master equations for different truncation orders $N$. We observe even slightly better agreement with the thermal populations when compared to Fig.~\ref{fig:thermaliz}, especially for the local Lindblad eq. for $N=1$. In Fig.~\ref{fig:coher-local}(c),(d) we plot the local magnitization currents as defined in the main text for site $i=1$ (c) and $i=2$ (d). We observe that the local Redfield eq. faithfully reproduces these currents. This is also true for the local Lindblad eq., apart from slight violations at the edges of the system.
 In a recent work \cite{TupkaryEtAl23} three criteria for a consistent Markovian master equation were formulated:  It obeys 1) complete positivity, 2) local conservation laws and 3) shows thermalization. While the standard Redfield eq.~violates criterion 1, the quantum optical ME violates criterion 2, and traditional local Lindblad eqs.~violate criterion 3. The local Lindblad master eq.~that we propose here fulfills all three criteria, however at the expense of possible approximation errors due to the truncation. Interestingly in Ref.~\cite{TupkaryEtAl23}, to find such an equation, the authors rely on a numerical optimization scheme. They find  that such an equation only exists if the corresponding Lindblad jump operators are acting over at least two sites. Consequentially, our work can provide an analytical derivation of the master equation that was targeted in that work numerically. Similar ideas were also used in Ref.~\cite{Shiraishi24}.

\begin{figure}
\centering
\includegraphics[scale=0.8]{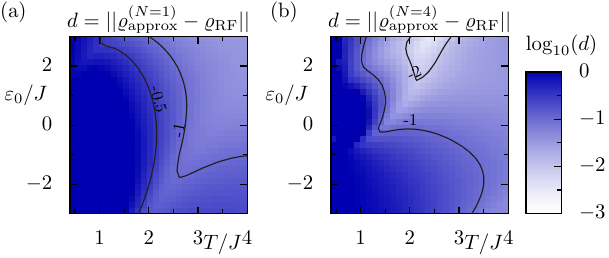}
\caption{Like in Fig.~\ref{fig:supp-rf-ord}, Fig.~\ref{fig:supp-rf-ord-drude-30} but for Lorentz-Drude bath with $E_\mathrm{D}=10h$. We use $L=4$.}
\label{fig:supp-rf-ord-drude-10}
\end{figure}

\subsection{Drude bath}

We show that the method also works well for bath models with Lorentz-Drude spectral density
\begin{align}
	J_\mathrm{D}(E) = \frac{\gamma E}{1+E^2/E_\mathrm{D}^2}
\end{align}
with Drude cutoff energy $E_\mathrm{D}$.

 \begin{figure}
\centering
\includegraphics[scale=0.8]{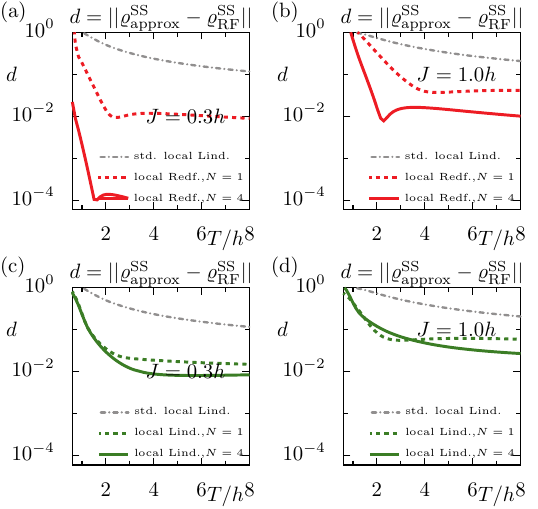}
\caption{Similar to Fig.~\ref{fig:dist-multiple} for  a Lorentz-Drude bath with $E_\mathrm{D}=10h$.} 
\label{fig:dist-multiple-drude-10}
\end{figure}
 \begin{figure}
\centering
\includegraphics[scale=0.8]{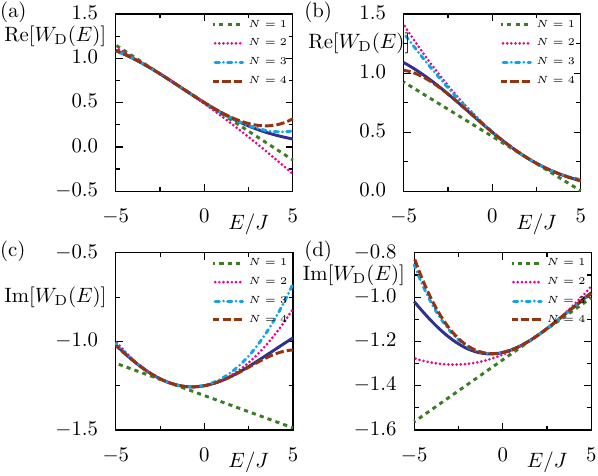}
\caption{Convergence of the Taylor series for real and imaginary part of the function $W_\mathrm{D}(E)$ for Lorentz-Drude bath with $E_\mathrm{D}=10J$ at $T=2J$ for expansion (a),(c)~around $\varepsilon_0=-2J$, (b),(d)~around $\varepsilon_0=2J$. We assume $J=h$.}
\label{fig:Wfunc-drude-10}
\end{figure}

  \begin{figure}
\centering
\includegraphics[scale=0.8]{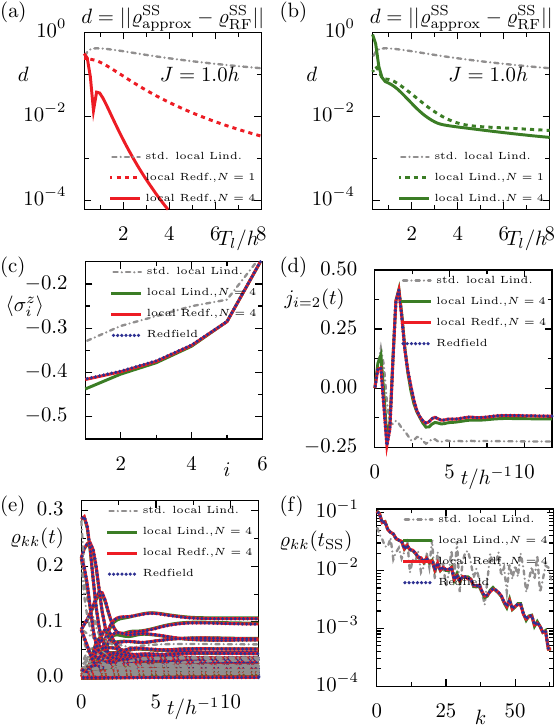}
\caption{Similar plots to the main text but for a nonequilibrium setup where $T_r=3T_l$, $T_l=2.2h$. (a) Like Fig.~2(d), (b) Like Fig.~3(a), (c) Like Fig.~3(b), (d) Like Fig.~3(c), (e) Like Fig.~2(a), (f) Like Fig.~2(b) of the main text.}
\label{fig:noneq}
\end{figure}

The corresponding bath-correlation function reads \cite{BreuerPetruccione,Schnell2019PHD,Thingna2013PHD}
\begin{align}
	&C(t) =  \frac{1}{\pi} \int_0^\infty  \mathrm{d}E J_\mathrm{D}(E) \left[ \mathrm{e}^{i E t} n_\mathrm{B}(E)+ \mathrm{e}^{-i E t} (1+n_\mathrm{B}(E))\right]\\
	& =\frac{\gamma E_\mathrm{D}^2}{2} \left[\cot\left(\frac{E_\mathrm{D}}{2T}\right) - i \right] \mathrm{e}^{-E_\mathrm{D}t}  -  \sum_{l=1}^\infty \frac{2 \gamma T \nu_l \mathrm{e}^{-\nu_l t}}{1-(\nu_l/E_\mathrm{D})^2}.
	\label{eq:C-exact}
\end{align}
for $t>0$. Here we use the bath occupation function $n_\mathrm{B}(E)=[\mathrm{e}^{E/T}-1]^{-1}$  and the  Matsubara frequencies $\nu_l = 2\pi T l$.

Using this, one finds \cite{Thingna2013PHD}
\begin{align}
	  W_\mathrm{D}(E)=   \int_0^\infty& \mathrm{d}\tau \mathrm{e}^{-i E \tau} C(\tau)\\
	  \begin{split}
	  = J_\mathrm{D}(E) n_\mathrm{B}&(E) + i  \sum_{l=1}^\infty \frac{2 \gamma T E  \nu_l }{[1-(\nu_l/E_\mathrm{D})^2](\nu_l^2 + E^2)}\\
	 &- i \frac{\gamma E_\mathrm{D}^2 E}{2(E_\mathrm{D}^2+E^2)}  \left[\cot\left(\frac{E_\mathrm{D}}{2T}\right) + \frac{E_\mathrm{D}}{E} \right].
	 \end{split}
	\label{eq:W-redf}
\end{align}
As in Fig.~\ref{fig:supp-rf-ord} we again benchmark the ad-hoc local Redfield equation for the XXZ spin chain in Fig.~\ref{fig:supp-rf-ord-drude-30} for a Lorentz-Drude bath with $E_\mathrm{D} = 30h$ and in Fig.~\ref{fig:dist-multiple-drude-30} for the local Redfield and local Lindblad and observe similar
good performance as in the case of the pure ohmic bath in the main text.  If one would neglect the imaginary part of $W_\mathrm{D}(E)$, this is somewhat expected, since the real part converges to the result of the pure ohmic spectral density for $E_\mathrm{D} \to \infty$. The imaginary part however is 
nontrivial and large, $\mathrm{Im} W_\mathrm{D}(0)= - \gamma E_D /2 = -3.75h$, since we do not add a renormalization term to the total Hamiltonian as for example in the Caldeira-Leggett model \cite{Thingna2013PHD}.

In Fig.~\ref{fig:supp-rf-ord-drude-10} and Fig.~\ref{fig:dist-multiple-drude-10}, we show the same plots for a lower value of the   Drude cutoff, $E_\mathrm{D}=10h$. We observe a much slower convergence than in the pure ohmic and the high-cutoff case. This can be somewhat understood from Fig.~\ref{fig:Wfunc-drude-10}, where we observe also slow convergence of the Taylor series due to the more complex curvature of the function at lower values of the cutoff.

\subsection{Nonequilibrium steady states}

In the main text we restrict ourselves to equilibrium steady states, i.e.~$T_l=T_r$. Nevertheless, the local Redfield- and local Lindblad master equations can also be used to describe proper nonequilibrium configurations that feature nonequilibrium steady states. In our model, this can be achieved by setting $T_l \neq T_r$, leading to a steady state with a finite heat current through the system. In Fig.~\ref{fig:noneq} we show similar plots as for the equilibrium case for a temperature of  the right bath that is three times of the one of left bath, $T_r=3T_l$. As we observe in Fig.~\ref{fig:noneq}(a) the error of the local Redfield equation is slightly lower than at equilibrium at the same temperature $T_l$, which suggests that the error is dominated by the error from the hotter bath $T_r$. As observed in Fig.~\ref{fig:noneq}(c),(d) the local Redfield- and Lindblad master equations successfully describe the steady state magnetization profile as well as the dynamics and the nonzero steady value of the local magnetization currents.

\subsection{Numerical details on quantum trajectories}

For the quantum trajectories we refer to the original algorithm of Ref.~\cite{KMolmer1993}.
No symmetry arguments are used, the full Hilbert space vectors are regarded. 
We only rely on the sparsity of all operators as discussed in the main text.
The trajectories are calculated on a small cluster of 10 nodes, 10th-13th gen.\ Intel Core i9 CPUs. Average memory
requirement is 10GB per job. Typical runtime for 50 trajectories are 30 hours. Overall the results
for 5000 trajectories were obtained after ca.~3 days.

\bibliography{mybib}